\pdfoutput=1
\newcommand*{\ATLASLATEXPATH}{latex/}
\documentclass[cernpreprint,texlive=2016,txfonts,UKenglish]{latex/atlasdoc}

\usepackage{\ATLASLATEXPATH atlaspackage}
\usepackage{amsmath}
\usepackage{siunitx}
\usepackage{graphicx}
\usepackage{\ATLASLATEXPATH atlasbiblatex}

\usepackage{\ATLASLATEXPATH atlascontribute}

\usepackage{\ATLASLATEXPATH atlasphysics}

\graphicspath{{logos/}{figures/}}

\usepackage{VLQ_Wb_2016_paper-defs}
\usepackage{latex/atlasjournal}
\usepackage{rotating}
\usepackage{verbatim}

\hypersetup{pdftitle={ATLAS document},pdfauthor={The ATLAS Collaboration}}

\graphicspath{{logos/}}

\AtlasTitle{Search for pair production of heavy vector-like quarks decaying to high-$\pt$ $W$ bosons and $b$ quarks in the lepton-plus-jets final state in $pp$ collisions at $\sqrt{s}$=13 TeV with the ATLAS detector}

\PreprintIdNumber{CERN-EP-2017-094}

\AtlasJournal{JHEP}
\AtlasJournalRef{JHEP 10 (2017) 141}
\AtlasDOI{https://doi.org/10.1007/JHEP10(2017)141}

\AtlasAbstract{%
A search is presented for the pair production of heavy vector-like
$T$ quarks, primarily targeting the $T$ quark decays to a $W$ boson and a $b$-quark.
The search is based on $36.1$~fb$^{-1}$ of $pp$ collisions at $\sqrt{s}$ = 13 TeV recorded in 2015 and 2016 with the ATLAS 
detector at the CERN Large Hadron Collider. 
Data are analysed in the lepton-plus-jets final state, including at least one $b$-tagged jet and a large-radius jet identified as originating from the hadronic decay of a high-momentum $W$ boson.
No significant deviation from the Standard Model expectation is observed in the reconstructed $T$ mass distribution.
The observed 95\% confidence level lower limit on the $T$ mass are 1350\,\GeV{} assuming 100\% branching ratio to $Wb$. In the SU(2) singlet scenario, the lower mass limit is 1170\,\GeV.
This search is also sensitive to a heavy vector-like $B$ quark decaying to $Wt$ and other final states. The results are thus reinterpreted to provide a 95\% confidence level lower limit on the $B$ quark mass at 1250\,\GeV{} assuming 100\% branching ratio to $Wt$; in the SU(2) singlet scenario, the limit is 1080\,\GeV{}. 
Mass limits on both $T$ and $B$ production are also set as a function of the decay branching ratios. 
The 100\% branching ratio limits are found to be applicable to heavy vector-like $Y$ and $X$ production that decay to $Wb$ and $Wt$, respectively.
}

\begin{document}


\maketitle

\tableofcontents

\clearpage

\section{Introduction}
\label{sec:intro}
The discovery of the Higgs boson by the ATLAS and CMS collaborations
is a major milestone in high-energy
physics~\cite{HIGG-2012-27,CMS-HIG-12-028}.  However, the underlying
nature of electroweak symmetry breaking remains unknown. Naturalness
arguments~\cite{naturalness} require that quadratic divergences arising from radiative
corrections to the Higgs boson mass are cancelled by a new
mechanism to avoid fine-tuning.  
This paper presents a search for pair production of vector-like quarks (VLQs) decaying into
third-generation quarks using the $pp$ collision data collected at the
Large Hadron Collider (LHC) in 2015 and 2016 at a centre-of-mass
energy of 13\,TeV.

Several new mechanisms have been proposed in theories beyond the Standard Model (BSM). In
supersymmetry, the cancellation comes from assigning superpartners to
the Standard Model (SM) bosons and fermions.  Alternatively, Little
Higgs~\cite{ArkaniHamed:2002qy,Schmaltz:2005ky} and Composite
Higgs~\cite{Kaplan:1983sm,Agashe:2004rs} models introduce a
spontaneously broken global symmetry, with the Higgs boson emerging as
a pseudo Nambu--Goldstone boson~\cite{Hill:2002ap}.  These latter
models predict the existence of VLQs, defined as
colour-triplet spin-1/2 fermions whose left- and right-handed chiral
components have the same transformation properties under the
weak-isospin SU(2) gauge
group~\cite{delAguila:1982fs,AguilarSaavedra:2009es}. 
Depending on the model, vector-like quarks are produced in SU(2) singlets, doublets or triplets of flavours $T$, $B$, $X$ or $Y$, in which the first two have the same charge as the SM top and $b$ quarks while the vector-like $Y$ and $X$ quarks have charge\footnote{All charges are quoted in units of $e$.} -4/3 and 5/3.
In addition, in these models, VLQs are expected to couple preferentially to third-generation
quarks~\cite{delAguila:1982fs,Aguilar-Saavedra:2013wba} and can have
flavour-changing neutral-current decays in addition to the
charged-current decays characteristic of chiral quarks. As a result,
an up-type $T$ quark can decay not only to a $W$
boson and a $b$ quark, but also to a $Z$ or Higgs boson and a
top quark ($T \to Wb$, $Zt$, and $Ht$). Similarly, a down-type $B$ quark can decay to a $Z$ or Higgs boson and a
$b$ quark, in addition to decaying to a $W$ boson and a top quark
($B \to Wt$, $Zb$, and $Hb$).  Instead, due to their charge, vector-like $Y$ quarks decay exclusively to $Wb$ while vector-like $X$ quarks decay exclusively to $Wt$.  To be consistent with the
results from precision electroweak measurements a small mass-splitting
between VLQs belonging to the same SU(2) multiplet is required, but no
requirement is placed on which member of the doublet is
heavier~\cite{Aguilar-Saavedra:2013qpa}.  Cascade decays such as
$T \to WB\to WWt$ are thus assumed to be kinematically forbidden.
Decays of VLQs into final states with first and second generation quarks,
although not favoured, are not
excluded~\cite{Atre:2008iu,Atre:2011ae}.

This search targets the $T\to Wb$ decay mode, although it
is sensitive to a wide range of branching ratios to the other two
decay modes as well as to vector-like $B$, $X$ and $Y$ production.
Previous searches in this decay mode by the ATLAS and CMS
collaborations did not observe a significant deviation from the SM
predictions. Those searches excluded VLQ masses below 740\,\GeV{} for any
combination of branching ratios and below 920\,\GeV{} for the
assumption of \BR$(T\to Wb)=1$~\cite{Aad:2015kqa,Chatrchyan:2013uxa}.
A recent search by the ATLAS collaboration at $\sqrt{s}=13$\,\TeV{} sets a lower limit of 1160\,GeV on the vector-like $T$ quark mass for the pure $Zt$ mode~\cite{Zt_13Tev}.

The event selection is optimised for \TTbar{} production with
subsequent decay to two high-\pt{} $W$ bosons and two $b$-quarks,
where one of the $W$ bosons decays leptonically and the other decays
hadronically.  To suppress the SM background, boosted jet
reconstruction
techniques~\cite{ATL-PHYS-PUB-2015-033,ATL-PHYS-PUB-2015-053} are used
to improve the identification of high-\pt{} $W$ bosons decaying
hadronically while rejecting events with hadronically decaying,
high-\pt{} top-quarks.  

The \TTbar{} system is reconstructed and the mass of the semi-leptonically
decaying VLQ candidate is used to discriminate between SM and VLQ
events.  Finally, a profile likelihood fit is used to test for the
presence of a VLQ signal as a function of $T$ and $B$ quark masses and
decay branching ratios.
The results are found to be equally applicable to either singlet or doublet weak-isospin configurations as well as applicable to the decays of $X$ and $Y$.


\section{ATLAS detector}
\label{sec:detector}
%
The ATLAS detector~\cite{PERF-2007-01} at the LHC is a multipurpose particle detector with a
forward-backward symmetric cylindrical geometry that covers nearly the entire solid angle 
around the collision point. It consists of an inner detector
surrounded by a thin superconducting solenoid providing a 2~T axial magnetic
field, electromagnetic and hadronic calorimeters, and a muon spectrometer.
The inner detector covers the pseudorapidity range\footnote{
The ATLAS Collaboration uses a right-handed coordinate system with its origin at the nominal
interaction point (IP) in the centre of the detector
and the $z$-axis along the beam pipe. The $x$-axis points from the IP to the
centre of the LHC ring, and the $y$-axis points upwards. Cylindrical coordinates
$(r,\phi)$ are used in the transverse plane, $\phi$ being the azimuthal angle
around the beam pipe. The pseudorapidity is defined in terms of the polar angle
$\theta$ as $\eta=-\ln\tan(\theta/2)$. Angular distance
is  measured in units of $\Delta R \equiv  \sqrt{(\Delta\eta)^2 + (\Delta\phi)^2} $.}
 $|\eta| < 2.5$. It consists of a silicon pixel detector, including the insertable 
B-layer installed after Run 1 of the LHC~\cite{IBL_1,IBL_2}, and a silicon microstrip detector surrounding the pixel detector, 
followed by a transition radiation straw-tube tracker.
Lead/liquid-argon sampling calorimeters provide electromagnetic
energy measurements with high granularity and a hadronic (steel/scintillator-tile)
calorimeter covers the central pseudorapidity range ($|\eta| < 1.7$).
The end-cap and forward regions are instrumented with liquid-argon calorimeters for
both the electromagnetic and hadronic energy measurements up to $|\eta| = 4.9$.
The outer part of the detector consists of a muon spectrometer with high-precision tracking chambers for coverage up to $|\eta|$ = 2.7, fast detectors for triggering over $|\eta|$ < 2.4, and three large superconducting toroid magnets with eight coils each.
The ATLAS detector has a two-level trigger system to 
select events for offline analysis~\cite{Aaboud:2016leb}.


\section{Data and simulation}
\label{sec:Samples}
This search utilises a data set corresponding to $36.1\pm1.2$~\ifb{} of integrated luminosity from $pp$ collisions at $\sqrt{s}=13$\,\TeV{} collected by the ATLAS experiment, with $3.2$~\ifb{} collected in 2015 and $32.9$~\ifb{} collected in 2016~\cite{lumi8tev}.
Data are only used if all ATLAS detector subsystems were operational. 
In all simulated events used in this search, the top quark and Higgs boson masses were set to 172.5\,\GeV{} and
125\,\GeV, respectively.

Simulated \TTbar{} events were generated with the leading-order (LO) generator \textsc{Protos} v2.2~\cite{protos} using the NNPDF2.3 LO parton distribution function (PDF) 
set and a set of tuned parameters called the A14 tune~\cite{A14} for the underlying-event description and passed to \textsc{Pythia 8.186}~\cite{Pythia81} for parton showering and fragmentation. 
The samples were generated for an SU(2) singlet $T$ VLQ, but with equal branching ratios of the $T$ quark 
to each final state.
To check the dependence of the results on the weak-isospin of the VLQ, one sample was also generated using the SU(2) doublet model including only the $T$ contributions.
The signal samples are normalised to pair-production cross-sections computed using
\textsc{Top++ v2.0}~\cite{topplusplus}, including next-to-next-to-leading-order (NNLO) quantum chromodynamics (QCD) corrections and soft-gluon resummation to 
NNLL accuracy~\cite{Cacciari:2011hy,Beneke:2011mq,Baernreuther:2012ws,Czakon:2012zr,Czakon:2012pz,Czakon:2013goa},  
and using the MSTW 2008 NNLO PDF set. 
Their cross-sections vary from $3.38\pm0.25$~pb ($m_{T}=500$\,\GeV) to $3.50\pm0.43$~fb ($m_{T}=1400$\,\GeV). 
Theoretical uncertainties are evaluated from variations of the factorisation and renormalisation scales, as well as from uncertainties in the PDFs and 
$\alphas$. The latter two represent the largest contribution to the overall theoretical uncertainty in the signal cross-sections and are calculated 
using the PDF4LHC~\cite{pdf4lhc} prescription with the MSTW 2008 68\% CL NNLO, CT10 NNLO~\cite{pdf1,pdf2} and NNPDF2.3~\cite{pdf3} 5f FFN PDF sets. 
Two benchmark signal scenarios are considered, along with a full scan of the branching-ratio plane.  The first benchmark corresponds to a $T$ quark that decays 100\% to $Wb$ and the second corresponds to the SU(2) singlet $T$ quark scenario, which predicts branching ratios of $\sim$50\%, $\sim$25\%, $\sim$25\% to $Wb$, $Zt$ and $Ht$, respectively~\cite{Aguilar-Saavedra:2013qpa}.
Samples were also generated for \BBbar\ production for the reinterpretation of this search. They were produced using the same generator and normalised in the same way as \TTbar.  As with \TTbar, two benchmark signal scenarios are considered, along with a full scan of the branching-ratio plane.  The first benchmark assumes \BR($B\to Wt)=1$ -- which also corresponds to the SU(2) ($B$,$T$) doublet hypothesis -- and the second corresponds to the SU(2) singlet $B$ quark scenario, which predicts branching ratios of $\sim$50\%, $\sim$25\%, $\sim$25\% to $Wt$, $Zb$ and $Hb$, respectively~\cite{Aguilar-Saavedra:2013qpa}.

The main SM backgrounds that are studied using simulated samples are due to \ttbar{}, \wjets{}, \zjets{}, diboson, single top quark, and \ttbar{}$+V$ ($V$ = $W$,$Z$) production.  
The multi-jet background is estimated using a data-driven technique discussed in Section \ref{sec:Backgrounds}. 
The nominal \ttbar{} MC sample was generated with \textsc{Powheg-Box v2} interfaced with \textsc{Pythia 6.428}~\cite{Powheg,Pythia6} 
for the parton shower and hadronisation, using the \textsc{Perugia2012} tune~\cite{Perugia} and the 
CT10 PDF set, and setting the $h_{\mathrm{damp}}$ parameter to the mass of the top quark. 
To estimate \ttbar{} modelling uncertainties, described in Section~\ref{sec:systematic_modelling}, additional samples were generated 
using \textsc{Powheg-Box v2} interfaced with \textsc{Herwig++ 2.7.1}~\cite{Herwig}, \textsc{Powheg-Box v2} interfaced with \textsc{Pythia 8.186}, and \textsc{MG5\_aMC@NLO 2.1.1} 
interfaced with \textsc{Pythia 8.186}~\cite{Madgraph}.  Further, samples with \textsc{Powheg-Box v2} interfaced with \textsc{Pythia 6.428} were generated 
varying the factorisation and normalisation scales by 2 and 0.5, as well as the next-to-leading-order (NLO) radiation factor, $h_{\mathrm{damp}}$, between $m_{\text{top}}$ and twice $m_{\text{top}}$. 
The \ttbar{} samples are normalised to the NNLO cross-section, including NNLO QCD corrections and soft-gluon resummation to NNLL accuracy, as done for the signal samples.   

Single top quark production (called `single top' in the following) in the $Wt$- and $s$-channels was also generated with \textsc{Powheg-Box v2} interfaced with \textsc{Pythia 6.428}, 
while single top production in the $t$-channel was generated with \textsc{Powheg-Box v1} interfaced with \textsc{Pythia 6.428} for the parton shower and hadronisation. 
Single-top samples were generated using the \textsc{Perugia2012} tune and the CT10 PDF set.  The single top 
cross-sections for the $t$- and $s$-channels are normalised to their next-to-leading-order (NLO) predictions, while for the $Wt$-channel the cross-section is 
normalised to its NLO+NNLL prediction~\cite{Kidonakis:2011wy}. For \wjets{}, \zjets{}, and diboson ($WW$, $WZ$, $ZZ$) samples, the \textsc{Sherpa 2.2.1} generator~\cite{Sherpa} 
was used with the CT10 PDF set. The \wjets{} and \zjets{} production samples are normalised to the NNLO cross-sections~\cite{Anastasiou:2003ds,Vjets_xs,Gavin:2010az}. For diboson production, 
the generator cross-sections (already at NLO) are used for sample normalisation. 
The \ttbar{}$+V$ background is modelled using samples produced with \textsc{MG5\_aMC@NLO 2.1.1} interfaced with \textsc{Pythia 8.186}, using the A14 tune and the NNPDF2.3 LO PDF set. The \ttbar{}+$V$ samples are normalised to their
respective NLO cross-sections~\cite{Madgraph}.  

All simulated samples were produced using the ATLAS simulation infrastructure~\cite{ATLASSIM}, 
using the full \textsc{GEANT4}~\cite{Geant4} simulation of the ATLAS detector and reconstructed with the same software as used for the data.
Multiple overlaid proton--proton collisions in the same or nearby bunch crossings (pile-up) were simulated at rates matching that of the data; they were 
modelled as low $\pt$ multi-jet production using the \textsc{Pythia 8.186} generator and tune \textsc{A2}~\cite{ATLAS:2012uec}.


\section{Analysis object selection}
\label{sec:ObjectDefinitions}
Reconstructed objects are defined by combining information from different detector subsystems.  This section outlines the criteria used to identify and select the reconstructed objects used in the analysis. Events are required to have at least one vertex candidate with at least two tracks with $\pt >$ 500\,MeV. The primary vertex is taken to be the vertex candidate with the largest sum of squared transverse momenta of all associated tracks.

To reconstruct jets, three-dimensional energy clusters in the calorimeter, assumed to represent massless particles coming from the primary vertex, are grouped together using the anti-$k_t$ clustering algorithm~\cite{antikt1,antikt2, Cacciari:2011ma} with a radius parameter of 0.4 (1.0) for small-$R$ (large-$R$) jets.  Small-$R$ jets and large-$R$ jets are clustered independently.

Small-$R$ jets are calibrated using an energy- and $\eta$-dependent calibration scheme, with \textit{in situ} corrections based on data~\cite{JetCalib}, and
are selected if they have $\pt >$ 25\,\GeV{} and $|\eta| <$ 2.5. A multivariate jet vertex tagger (JVT) selectively removes small-$R$ jets that are identified as having originated from pile-up collisions rather than the hard scatter~\cite{jvt}. Jets containing $b$-hadrons are identified via an algorithm that uses multivariate techniques to combine information from the impact parameters of displaced tracks as well as topological
properties of secondary and tertiary decay vertices reconstructed within the jet.
A jet is considered $b$-tagged if the value for the multivariate discriminant is above
the threshold corresponding to an efficiency of 77\% for tagging a $b$-quark-initiated jet. The corresponding light-jet rejection factor is $\sim$130 and the charm-jet rejection factor is $\sim$6, as determined for jets with $\pt{}>20$\,\GeV{} and $|\eta|<2.5$ in simulated \ttbar{} events.

Large-$R$ jets are built using 
the energy clusters in the calorimeter~\cite{Aad:2016upy,ATLAS-CONF-2016-035} and then trimmed~\cite{Krohn2010} to mitigate the 
effects of contamination from multiple interactions and improve background rejection.  The jet energy and pseudorapidity are 
further calibrated to account for residual detector effects using energy and pseudorapidity dependent calibration factors derived from simulation.
The $k_t$-based trimming algorithm reclusters the jet constituents into subjets with a
finer-grained resolution (the $R$-parameter for subjets is set
to  $R_{\text{sub}}=0.2$). Subjets that contribute less than 5\% to the \pt{} of the large-$R$ jets are discarded.
The properties (e.g. transverse momentum and invariant mass) of the jet are
recalculated using only the constituents of the remaining subjets.
Trimmed large-$R$ jets are only considered if they have $\pt{}>200$\,\GeV{} and $|\eta|<2.0$.
To identify large-$R$ jets that are likely to have originated from the hadronic decay of $W$ bosons ($W_\text{had}$) and not from the hadronic decay of top quarks or multi-jet background, jet substructure information is exploited using the ratio of the energy correlation functions $D_2^{\beta=1}$~\cite{Larkoski:2014gra,Larkoski:2015kga} and jet mass~\cite{ATLAS-CONF-2016-035}.
Selected large-$R$ jets must pass both the substructure and mass requirements of the 50\%-efficient $W$-tagging working point~\cite{ATL-PHYS-PUB-2015-033}. 
To reduce the
contribution from the $t\bar t$ background, the $W_\text{had}$ candidate
must not overlap any $b$-tagged small-$R$ jets within $\Delta R
< 1.0$.  If multiple large-$R$ jets satisfy the above requirements,
the one with a mass closest to the mass of the $W$ boson is
selected as the $W_\text{had}$ candidate.


Electrons are reconstructed from energy deposits in the electromagnetic calorimeter matched to inner detector tracks.  
Electron candidates are required to satisfy likelihood-based identification criteria~\cite{ATLAS-CONF-2016-024} and must have $\pt^\text{lep} > 30$\,\GeV{} and $|\eta|<2.47$.
Electron candidates in the transition region between the barrel and endcap electromagnetic calorimeters, $1.37 < |\eta| < 1.52$, are excluded from this analysis. 
A lepton isolation requirement is implemented by calculating the quantity 
$I_R = 
\sum_{\Delta R
(\mathrm{track},\text{lep})
<R_\mathrm{cut}} 
\pt^\mathrm{track}$, 
 where $R_\mathrm{cut}$ is the smaller of 10 GeV/$p_T^\text{lep}$ and 0.2; the track associated with the lepton is excluded from the calculation. The electron must satisfy $I_R < 0.06\cdot p_T^\text{lep}$.
Additionally, electrons are required to have a track satisfying $\frac{|d_0|}{\sigma_{d_0}} < 5$ and $|z_0 \sin\theta| < 0.5$\,mm, where $d_0$ is the transverse impact parameter and $z_0$ is the $r$--$\phi$ projection of the impact point onto the $z$-axis.
An overlap-removal procedure prevents double-counting of energy between an electron and nearby jets 
by removing jets if the separation between the electron and jet is within $\Delta R<$ 0.2 and removing electrons if the separation is within 0.2 $<\Delta R<$ 0.4.
In addition, a large-$R$ jet is removed if the separation between the electron and the large-$R$ jet is within $\Delta R < 1.0$.

Muons are reconstructed from an inner detector track matched to muon spectrometer tracks or track segments~\cite{Aad:2016jkr}. Candidate muons are required to pass quality specifications based on information from the muon spectrometer and inner detector.
Furthermore, muons are required to be isolated from detector activity using the same criterion that is applied to electrons and their associated tracks must satisfy $|z_0 \sin\theta| < 0.5$\,mm and $\frac{|d_0|}{\sigma_{d_0}} < 3$.
Muons are selected if they have $\pt{}>30$\,\GeV{} and $|\eta|<2.5$. 
An overlap-removal procedure is also applied to muons and jets.  If a muon and a jet with at least three tracks are separated by $\Delta R < \min(0.4, 0.04+10\,\GeV/\pt{}^{\mu})$ the muon is removed;  if the jet has fewer than three tracks, the jet is removed.

For a given reconstructed event, the magnitude of the negative vector
sum of the \pt{} of all reconstructed leptons and small-$R$ jets is defined as the missing transverse momentum (\met{})~\cite{met}.  
An extra term is included to account for `soft' energy from 
inner detector tracks that are not matched
to any of the selected objects but are consistent with originating from the primary vertex.

The four-momentum of the neutrino can be analytically determined in each event using the missing transverse momentum vector $\vec{E}_\mathrm{T}^\mathrm{miss}$ and assuming the lepton--neutrino system has an invariant mass equal to that of the $W$ boson. Nearly half 
of the events are found to produce two complex solutions. When complex solutions are obtained, a real solution is 
determined by minimising a $\chi^2$ parameter based on the difference between the mass of the lepton--neutrino system and the measured value of the $W$ 
boson mass. In the case of two real solutions, the solution with the smaller absolute value of the longitudinal momentum is used.


\section{Analysis strategy}
\label{sec:AnalysisStrategy}
This search targets the decay of pair-produced VLQs, $\TTbar{}$, where one $T$ quark decays to $Wb$ and the other 
decays to $Wb$, $Zt$ or $Ht$. Since previous searches from ATLAS and 
CMS have excluded VLQs decaying to $Wb$ at 95\% confidence level (CL) for masses below 920\,\GeV, this search focuses on the decays of higher-mass VLQs. 
The final state consists of a high-\pt{} 
charged lepton and missing transverse momentum from the decay of one of the $W$ bosons, a high-momentum large-$R$ jet from the hadronically decaying $W$ boson, and multiple $b$-tagged jets.
The event preselection is described in Section~\ref{sec:EventSelection} and the reconstruction of the \TTbar{} system is discussed  in Section~\ref{sec:tt_reco}. The classification of events into signal and control regions follows in Section~\ref{sec:ControlRegions}.

The search for the \BBbar\ signal uses the same selection criteria, with no further optimization.


\subsection{Event preselection}
\label{sec:EventSelection}
Events are required to pass a single-electron or single-muon trigger.  The 2015 data were collected using electron triggers with $\et$ thresholds of 24, 60, and 120\,\GeV.  The 2016 data were collected using electron triggers with $\et$ thresholds of 26, 60, and 140\,\GeV.
For the 2015 electron triggers, the highest-$\et$ trigger had a looser quality requirement on the trigger object than the triggers with lower $\et$ thresholds. 
For the 2016 electron triggers, the trigger with the lowest $\et$ threshold had stringent requirements on the quality of the trigger object, 
as well as requirements on its isolation from other activity in the detector.  
The highest and second highest $\et$ triggers had no requirement on isolation and had progressively looser 
quality requirements. 
Muon triggers with $\pt$ thresholds of 20 (26)~\GeV{} and requirements on isolation were used in 2015 (2016).
Additionally, a high-$\pt$ muon trigger with a threshold of 50\,\GeV{} and no isolation requirement was used in both 2015 and 2016 data.

In addition to the trigger requirement, events must have at least one primary vertex with at least two associated tracks.  
Exactly one lepton candidate (electron or muon), as described in Section~\ref{sec:ObjectDefinitions}, is required.
Signal events are expected to have a high jet multiplicity, since they include two $b$-jets as well as one jet from the hadronic decay of the $W$ boson. 
Therefore, at least three small-$R$ jets are required, of which at least one must be $b$-tagged. 
At least one boosted hadronic $W$ candidate is required and the \etmiss{} is required to be greater than 60\,\GeV.

After this selection, backgrounds with large contributions include \ttbar{}, \wjets{}, and single-top events.  
Other SM processes, including diboson, \zjets{},  \ttbarV{} and multi-jet production, make a smaller but non-negligible contribution; these small backgrounds are collectively referred to as `Others'.


\subsection{\TTbar{} reconstruction}
\label{sec:tt_reco}

After preselection, the four-momenta of the hadronic and
semi-leptonic VLQ candidates are reconstructed using the selected lepton
candidates, large-$R$ jets, small-$R$ jets, and missing transverse
momentum of the event.
VLQ candidates ($T\to Wb$) are formed by pairing each $W$ boson
candidate with a $b$-quark candidate.  
If there are two or more $b$-tagged jets in
the event, the two highest-\pt{} $b$-tagged jets are selected as the
$b$-quark candidates.  Both possible pairings of the $b$-quark
candidates with the $W_\text{had}$ and semi-leptonically decaying $W$ boson ($W_\text{lep}$) candidates are
tested and the pairing that minimises the absolute value of the mass
difference between the semi-leptonically and hadronically reconstructed VLQ candidates, $|\Delta m|$, is chosen.  If the event has only one $b$-tagged
jet, that jet is used as one of the $b$-quark candidates and then all
permutations with the remaining small-$R$ jets are tested to find the
configuration that minimises $|\Delta m|$.

The final discriminating variable used in the statistical analysis is $m_T^\text{lep}$, the reconstructed mass of the semi-leptonically decaying vector-like $T$ quark candidate. This is found to provide the best expected signal sensitivity.
Figure~\ref{fig:TTreco:signals} shows $m_T^\text{lep}$ for benchmark $T$ and $B$ quark signal models and \ttbar{} production in the signal region (defined in
Section~\ref{sec:baseline_selection}) after the reconstruction
algorithm is applied.  
The reconstructed masses for the signal and \ttbar{} background are shown to peak at the
generated $T$ and top-quark masses, respectively.  
The tails arise from misreconstructed $T$ candidates.
As expected, the reconstruction algorithm does not reconstruct the $B$ mass, yet the variable nonetheless provides separation power between the signal and the \ttbar{} background.

\begin{figure}[ht!]
\centering
\includegraphics[width=0.48\textwidth]{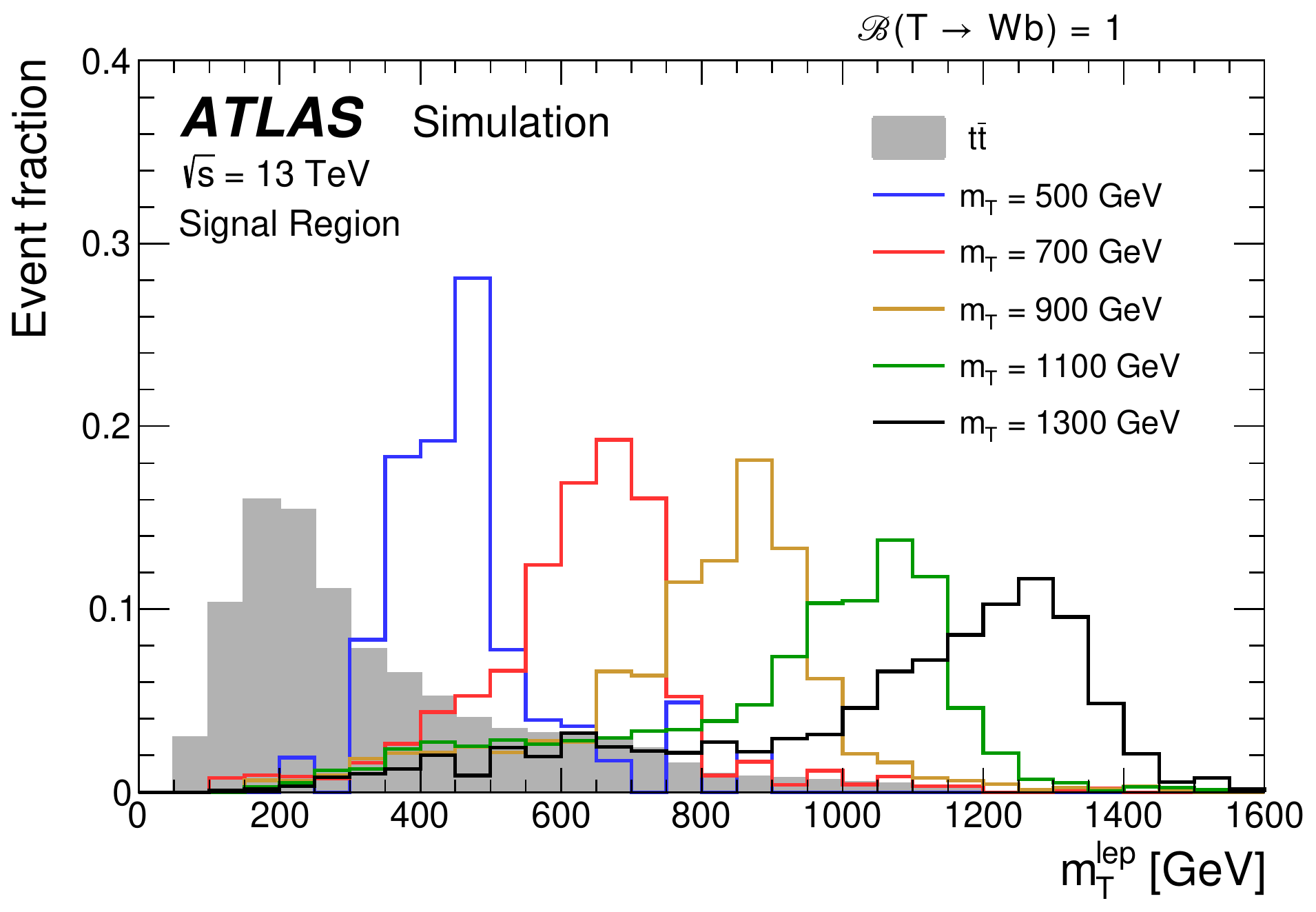}
\includegraphics[width=0.48\textwidth]{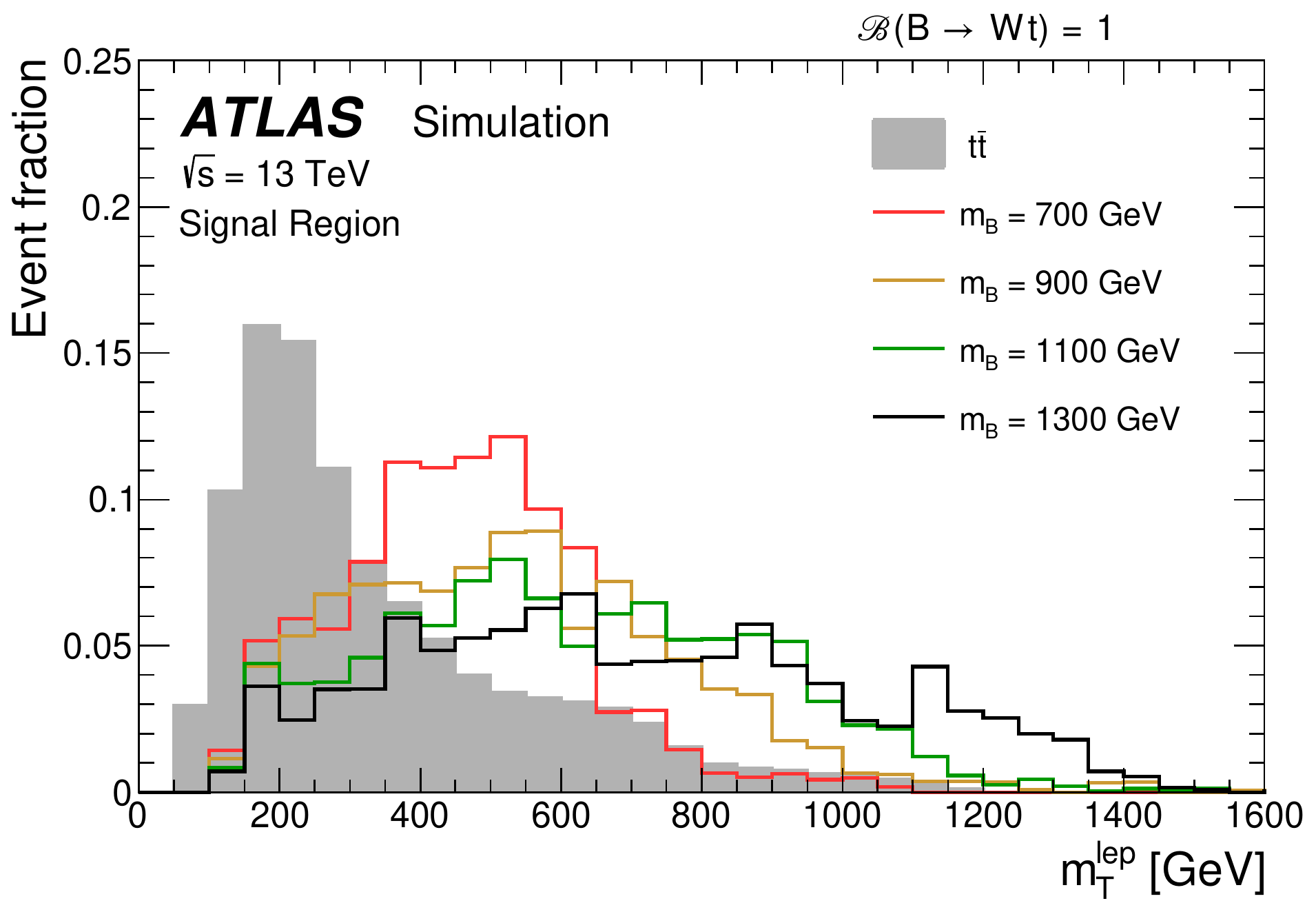}
\caption{The reconstructed leptonic $T$ quark mass in the signal region is shown for the \ttbar{} background and a few signal mass points, for the signal models \BR$(T \rightarrow W b)= 1$ (left) and for the signal models \BR$(B \rightarrow Wt) = 1$ (right). In both figures, the distributions are normalised to unity for comparison of the relative shapes at each mass point. Due to the limited Monte Carlo sample size, the \ttbar{} distribution has been smoothed.}
\label{fig:TTreco:signals}
\end{figure}

\subsection{Classification of event topologies}
\label{sec:ControlRegions}
A \ttbar{} control region
is used to constrain the production rate of \ttbar{} events as well as systematic uncertainties related to \ttbar{} modelling. 
The signal and control regions are described in detail in Section~\ref{sec:baseline_selection} and Section~\ref{sec:ttbarCR}.
The scalar sum of \met{} and the transverse momenta of the lepton and all small-$R$ jets, $S_\mathrm{T}$, and the separation between the lepton and neutrino, $\Delta R(\mathrm{lep}, \nu)$, are used to define the two regions. These regions are shown in Figure~\ref{fig:2Dplots} after applying the event pre-selection, and described below.

\begin{figure}[b!]
\centering
\includegraphics[width=0.48\textwidth]{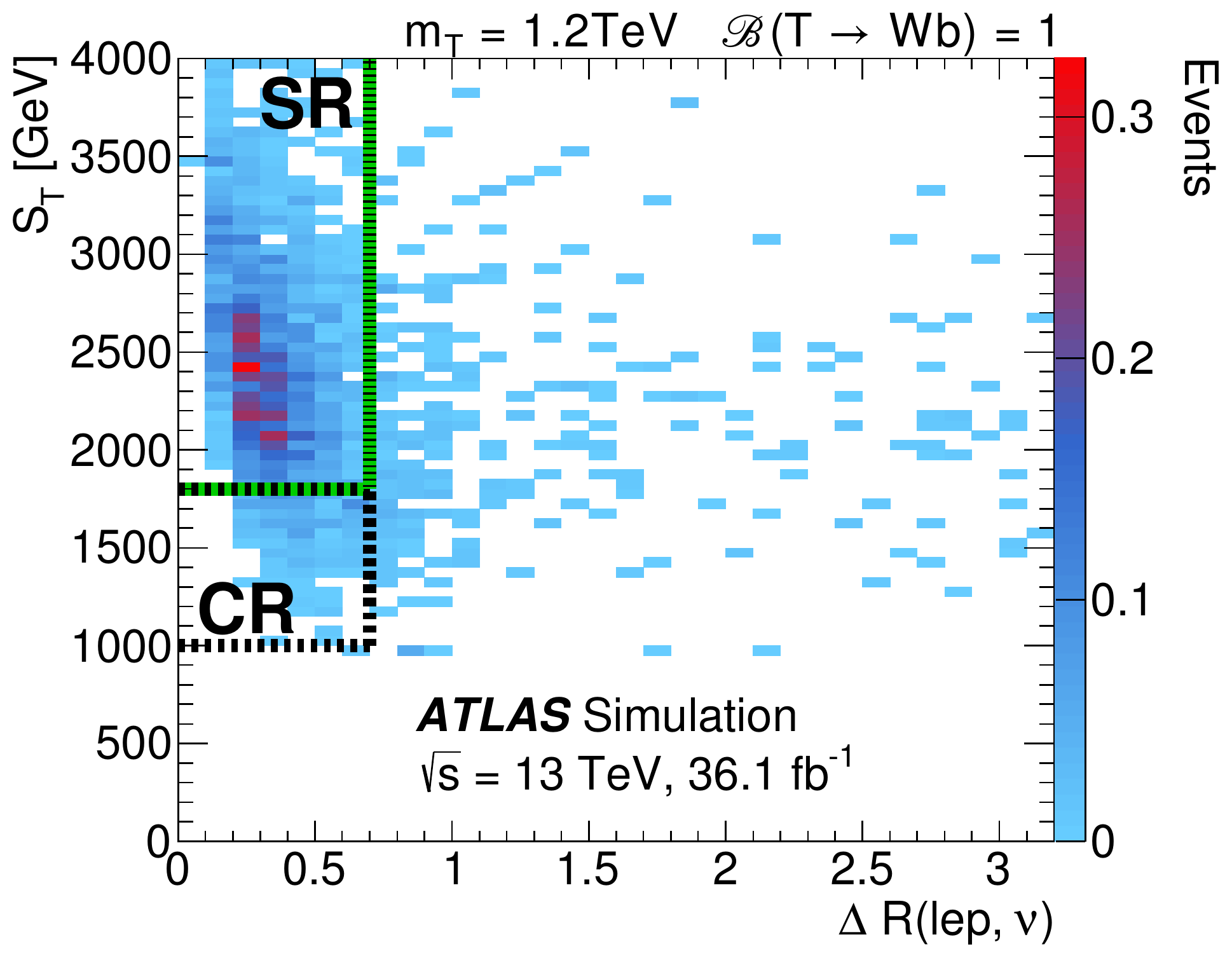}
\includegraphics[width=0.48\textwidth]{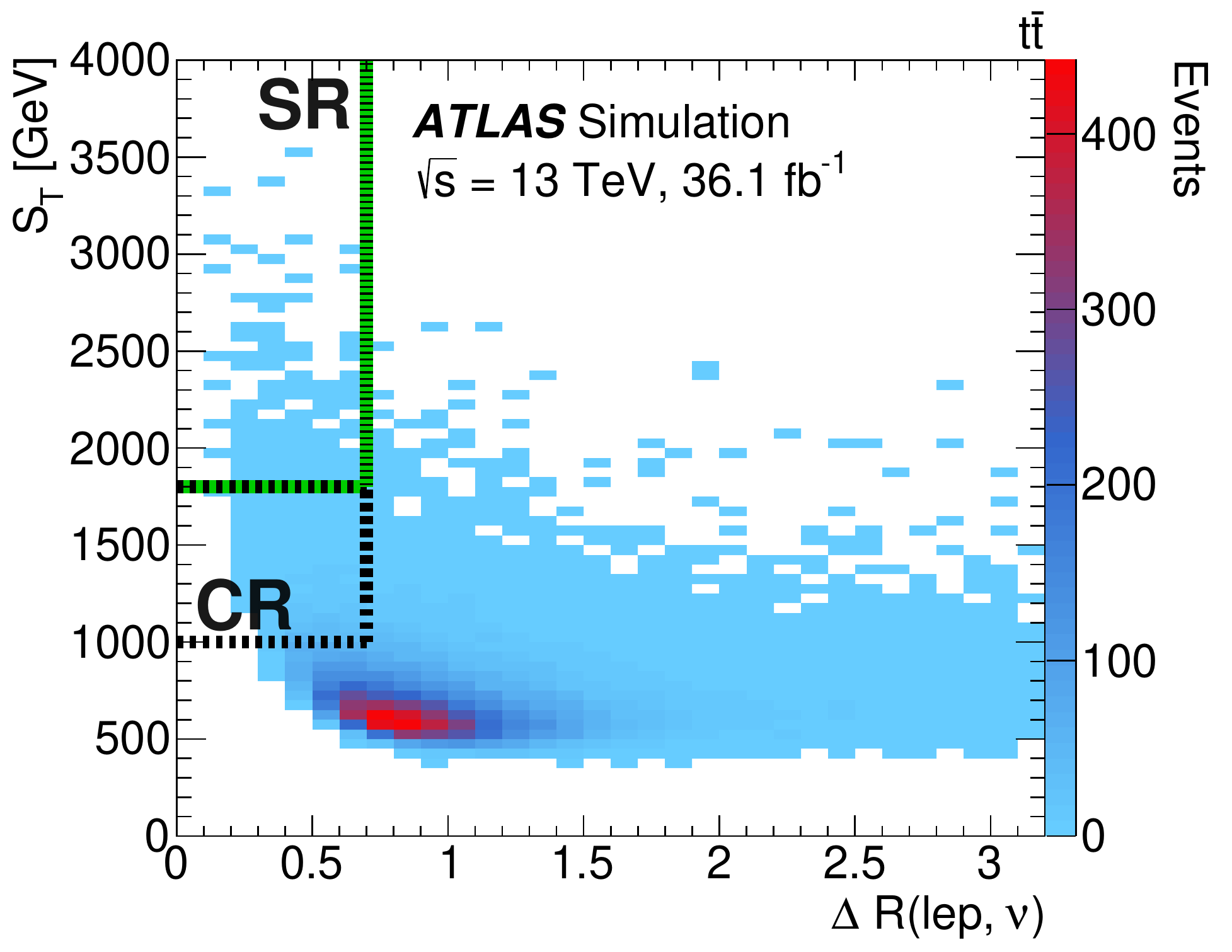}
\caption{The signal region (SR) and control region (CR) are shown in a two-dimensional plane of $S_\mathrm{T}$ and $\Delta R(\mathrm{lep}, \nu)$, overlaying the expected signal distribution for \BR($T \to Wb) = 1$ and a mass of 1.2\,\TeV{} (left) and overlaying the distribution of the dominant \ttbar{} background (right). }
\label{fig:2Dplots}
\end{figure}

\subsubsection{Signal region definition}
\label{sec:baseline_selection}
After the event pre-selection described in Section~\ref{sec:EventSelection}, further requirements are applied to reduce the contribution of SM backgrounds relative to signal. 
Events in the signal region are selected based on their characteristic boosted topology with a high-\pt{} $W$ boson and larger separation between the $W$ boson and the $b$ quarks. 
Events are required to have $\Delta R(\mathrm{lep},\nu) < 0.7$, arising from a boosted leptonically decaying $W$ boson. In addition, $S_T$ is required to be greater than 1800\,\GeV{}.
This requirement is found to maximise the expected sensitivity to VLQ masses above 1 TeV.
In order to reject both the \ttbar{} and single-top (mostly $Wt$-channel)  backgrounds, an additional requirement is put on the difference between the reconstructed masses of the leptonic and hadronic VLQ candidates, $ | \Delta m | = | m_T^\mathrm{had} - m_T^\mathrm{lep} | < 300$\,\GeV; this selection criterion is optimised to provide the best expected sensitivity. 

The expected numbers of events in the signal region for the background processes and signal hypothesis with mass $m_T$ = 1~\,TeV are shown in Table~\ref{tab:loose_yields}. For a signal model with \BR$(T\to Wb)=1$, the acceptance times efficiency of the full event selection ranges from 0.2\% to 4.0\% for VLQ masses from $m_T$ = 500 to 1400\,\GeV.  For the SU(2) singlet $T$ scenario, for which \BR$(T\to Wb)$ is approximately 50\% for the mass range of interest, the signal acceptance ranges from 0.1\% to 2.0\%.

\begin{table}[!ht]{}
\begin{center}
  \caption{Event yields for background sources and several signal models in the signal and control regions. 
  The yields are given before the profile likelihood fit described in Section~\ref{sec:Results}. The quoted uncertainties include statistical and systematic uncertainties; for the \ttbar{} background no cross-section uncertainty is included. The contributions from dibosons, $Z$+jets, $ttV$ and multi-jet production are included in the Others category.}
    \label{tab:loose_yields}
    \begin{tabular}{ l S[table-format=3.0, table-figures-uncertainty=1] S[table-format=3.0, table-figures-uncertainty=1] S[table-format=3.0, table-figures-uncertainty=1]} 
     \hline
     \hline
   \multicolumn{1}{l}{Sample} & \multicolumn{1}{l}{Signal region} & \multicolumn{1}{l}{Control region} \\
  \hline
  \ttbar{}       &  55 \pm  26  &  720 \pm 130  \\
   $W$+jets   &  9  \pm  4   &  78 \pm  41 \\
   Single top &  15 \pm  15  &  160 \pm 110 \\
  Others  &  12  \pm 10  &  82 \pm  66 \\
 \hline
 Total Background  &  91  \pm 35   &  1040 \pm 200\\
 \hline
 Signal ($m_T =$ 1~\TeV, \BR$(T \rightarrow Wb) = 1$) &   45 \pm 4  &  15 \pm 2 \\
 Signal ($m_T =$ 1~\TeV, SU(2) singlet) &   21 \pm 2  &  8 \pm 1  \\
 Signal ($m_B =$ 1~\TeV, \BR$(B \rightarrow Wt) = 1$) &   46 \pm 4  &  21 \pm 2 \\
 Signal ($m_B =$ 1~\TeV, SU(2) singlet) &   18 \pm 2  &  8 \pm 1  \\
 \hline
 Data  &  58   &  972 \\
 \hline
 \hline
 \end{tabular}
    \end{center}
  \end{table}

\subsubsection{Control region definition}
\label{sec:ttbarCR}

In this analysis, SM \ttbar{} production is the dominant background process.  To constrain the rate of \ttbar{} production in the signal region, as well as to constrain some uncertainties related to \ttbar{} modelling, a control region is included in the statistical analysis. This region is defined by only changing the requirement on $S_\text{T}$ to 1000 \GeV\,$ < S_\text{T} < $1800 \GeV.  This window is chosen to be as close as possible to the signal region, while still retaining a large number of background events.  Both the lower requirement on the control region and the requirement separating the signal and control regions were optimised to maximise the expected sensitivity to the signal with a mass of 1000\,\GeV{} and \BR($T\to Wb$) = 1.

\subsection{Multi-jet background estimation}
\label{sec:Backgrounds}
The multi-jet background originates from either the
misidentification of a jet as a lepton candidate (fake lepton) or from the presence of a
non-prompt lepton (e.g., from a semileptonic $b$- or $c$-hadron decay)
that passes the isolation requirement.  The multi-jet shape,
normalisation, and related systematic uncertainties are estimated from
data using the matrix method (MM)~\cite{Aad:2010ey}.  The MM
exploits the difference in efficiency for prompt leptons to pass loose and tight
quality requirements, obtained from $W$ and
$Z$ boson decays, and non-prompt or fake lepton candidates, from
the misidentification of photons or jets.
The efficiencies, measured in dedicated control regions, are
parameterised as functions of the lepton candidate $\pt$ and $\eta$,
$\Delta \phi$ between the lepton and jets, and the $b$-tagged jet
multiplicity. 

The event selection used in this analysis significantly reduces the contribution of the multi-jet background in the signal and control regions, to the point where statistical uncertainties make the MM prediction unreliable.
In order to obtain a reliable prediction, the requirements on $S_T$ and $\Delta R(\mathrm{lep},\nu)$ are released to 1200~\GeV\, and 1.5, respectively. In this region the MM prediction and the small Monte Carlo derived backgrounds (diboson, $Z$+jets and $ttV$) are studied and their shapes are found to be compatible.
This selection is thus used to determine the ratio of the multi-jet production to the small Monte Carlo derived backgrounds. The ratio is then assumed to be the same in the signal and control regions and is used to scale those small MC derived backgrounds in order to account for the additional contribution from multi-jet backgrounds. This scaling was found to be stable under small changes to the definition of the looser selection. In the signal region, the contribution from the multi-jet background to the total background is around 6\%.


\section{Systematic uncertainties}
\label{sec:Systematics}
The systematic uncertainties are broken down into four broad categories: luminosity and cross-section uncertainties, 
detector-related experimental uncertainties, uncertainties in data-driven background estimations, and modelling uncertainties in simulated background 
processes. Each source of uncertainty is treated as 
a nuisance parameter in the fit of the leptonic $T$ mass distribution, and shape effects are taken into account where relevant.
Due to the tight selection criteria applied, the analysis is limited by the statistical uncertainty; the systematic uncertainties only mildly degrade the sensitivity of the search.

\subsection{Luminosity and normalisation uncertainties}
\label{sec:systematic_Luminosity}

The uncertainty in the combined 2015+2016 integrated luminosity is 3.2\%. It is derived, following a methodology similar to that detailed in Ref.~\cite{lumi8tev}, from a preliminary calibration of the luminosity scale using $x$--$y$ beam-separation scans performed in August 2015 and May 2016. 
This systematic uncertainty is applied to all backgrounds and signal that are estimated using simulated 
Monte Carlo events, which are normalised to the measured integrated luminosity.

Theoretical cross-section uncertainties are applied to the relevant simulated samples.
The uncertainties for $W$/$Z$+jets and diboson production are 5\% and 6\%, respectively~\cite{Vjets_xs, dib_xs}. 
For the largest of these backgrounds, $W$+jets, a total uncertainty of 50\% in the normalisation is included. The pre-fit impact\footnote{The pre-fit effect on the signal strength parameter $\mu$ is calculated by fixing the corresponding uncertainty at $\theta\pm \sigma_{\theta}$, where $\theta$ is the initial value of the systematic uncertainty and $\sigma_{\theta}$ is its pre-fit uncertainty, and performing the fit again.
The difference between the default and the modified value of $\mu$, $\Delta\mu$, represents the effect on $\mu$ of this particular uncertainty (see section~\ref{Results:StatisticalMethod} for further details).} on the measured signal strength of the $W$+jets normalisation is less than 1\%. 
Two additional shape 
uncertainties are also considered, related to the heavy-flavour content in the $W$+jets background. 
These uncertainties are derived 
by varying each heavy-flavour component of the $W$+jets background individually by a factor of 1.5, while keeping the overall 
normalisation fixed. 
For single top production, the uncertainties are taken as 6\%~\cite{top_xs_1,top_xs_2}.
The normalisation of \ttbar{} is unconstrained in the fit.
For the data-driven multi-jet estimation, an uncertainty of 100\% is assigned to the normalisation, corresponding to the maximum range obtained by varying the values of the cuts on $S_T$ and $\Delta R(\mathrm{lep},\nu)$ when obtaining the multi-jet contribution to the `Others' background.  

\subsection{Detector-related uncertainties}
\label{sec:systematic_detector}
The dominant sources of detector-related uncertainties in the signal and background yields relate to the small-$R$ and large-$R$ jet energy scales and resolutions. 
The small-$R$ and large-$R$ jet energy scales and their uncertainties are derived by combining information from test-beam data, LHC collision data and simulation~\cite{ATL-PHYS-PUB-2015-015}.
In addition to energy scale and resolution uncertainties, there are also uncertainties in the large-$R$ mass and substructure scales and resolutions. 
These are evaluated similarly to the jet energy scale and resolution uncertainties and are propagated to the $W$-tagging efficiencies. At $\sim$2\%, the uncertainty in the jet energy resolution has the largest pre-fit impact on the measured signal strength, corresponding to a normalisation difference in the signal, \ttbar{}, and single top yields of 2\%, 2\%, and 14\%, respectively.

Other detector-related uncertainties come from lepton trigger efficiencies, identification efficiencies, energy scales and resolutions, the \met{} reconstruction, the $b$-tagging efficiency, and the JVT requirement. Uncertainties related to the efficiency for tagging $c$-jets have the largest pre-fit impact on the measured signal strength ($\sim$1\%). This originates from a change in normalisation of $\sim$3\% on both the signal and background yields.

\subsection{Generator modelling uncertainties}
\label{sec:systematic_modelling}

Modelling uncertainties are estimated for the dominant \ttbar{} and single-top backgrounds. 
The modelling uncertainties are estimated by comparing simulated samples with different configurations, described in Section~\ref{sec:Samples}.
The effects of extra initial and final state gluon radiation are estimated by comparing simulated samples generated with enhanced or reduced initial state radiation, changes to the $h_\text{damp}$ parameter, and different radiation tunes. This uncertainty has a 12\% normalisation impact on \ttbar{} in the signal region, resulting in a pre-fit impact of $\sim$1\% on the measured signal strength.
The uncertainty in the fragmentation, hadronisation and underlying-event modelling is estimated by comparing two different parton shower models, \textsc{Pythia} and \textsc{Herwig++}, while keeping the same hard-scatter matrix-element generator. This causes an 18\% shift in the normalisation of \ttbar{} in the signal region, resulting in a pre-fit impact of $\sim$3\% on the measured signal strength.
The uncertainty in the hard-scatter generation is estimated by comparing events generated with two different Monte Carlo generators, \textsc{MG5\_aMC@NLO} and \textsc{Powheg}, while keeping the same parton shower model. This uncertainty has a 38\% normalisation impact on \ttbar{} in the signal region, resulting in a pre-fit impact of only $\sim$4\% on the measured signal strength.

Modelling uncertainties in single top production are also included. In this analysis, $Wt$-channel production is the dominant contribution and the largest uncertainty comes from the method used to remove the overlap between NLO $Wt$ production and LO \ttbar{} production. The default method used is diagram removal, while the alternative method considered is diagram subtraction~\cite{DRDS}.  The full difference between the two methods is assigned as an uncertainty. This uncertainty has a 90\% normalisation impact on single top in the signal region resulting in a pre-fit impact of $\sim$5\% on the measured signal strength.


\section{Results}
\label{sec:Results}
\subsection{Statistical interpretation}\label{Results:StatisticalMethod}

The distribution of the reconstructed mass of the leptonically decaying $T$ quark candidate, $m_T^{\mathrm{lep}}$, in the signal and control regions is used to test for the presence of a signal. Hypothesis testing is performed using a modified frequentist method as implemented in \textsc{RooStats}~\cite{RooFit,RooFitManual} and based on a profile likelihood which takes into account the systematic uncertainties as nuisance parameters that are fitted to the data. 

The statistical analysis is based on a binned likelihood function $\mathcal{L}(\mu,\theta)$ constructed as a product of Poisson probability terms over all bins considered in the search. This function depends on the signal strength parameter $\mu$, a multiplicative factor to the theoretical signal production cross-section, and $\theta$, a set of nuisance parameters that encode the effect of systematic uncertainties in the signal and background expectations and are implemented in the likelihood function as Gaussian constraints. Uncertainties in each bin of the $m_T^{\mathrm{lep}}$ distributions due to finite size of the simulated samples are also taken into account via dedicated fit parameters and are propagated to $\mu$. 
In this analysis, the normalisation of the dominant \ttbar\ background is included as an unconstrained nuisance parameter; there are sufficient number of events in the control regions and low mass region of the signal region, where the signal contribution is small, to obtain a data-driven estimate of the \ttbar{} normalisation.
Nuisance parameters representing systematic uncertainties are only included in the likelihood if either of the following conditions are met: overall impact on the normalisation is larger than 1\%, or the shape of the uncertainty varies by more than 1\% between adjacent bins. This is done separately for each region and for each template (signal or background).
When the bin-by-bin statistical variation of a given uncertainty is significant, a smoothing algorithm is applied.

The expected number of events in a given bin depends on $\mu$ and $\theta$. The nuisance parameters $\theta$ adjust the expectations for signal and background according to the corresponding systematic uncertainties, and their fitted values correspond to the amounts that best fit the data. 
This procedure allows for a reduction of the impact of systematic uncertainties in the search sensitivity by taking advantage of the highly populated background-dominated control region (CR) included in the likelihood fit.

The test statistic $q_{\mu}$ is defined as the profile likelihood ratio,
$q_{\mu}=-2\mathrm{ln}(\mathcal{L}(\mu,\hat{\hat{\theta}}_{\mu})/ \mathcal{L}(\hat{\mu}, \hat{\theta}))$, 
where $\hat{\mu}$ and $\hat{\theta}$ are the values of the parameters that maximise the likelihood function (with the constraint 0$\leq  \hat{\mu} \leq \mu$), and $\hat{\hat{\theta}}_{\mu}$ are the values of the nuisance parameters that maximise the likelihood function for a given value of $\mu$. The compatibility of the observed data with the background-only hypothesis is tested by setting $\mu=0$ in the profile likelihood ratio: $q_{0}=-2\mathrm{ln}(\mathcal{L}(0,{\hat{\hat{\theta}}}_{0})/\mathcal{L}({\hat{\mu}},{\hat{\theta}}))$.
In the absence of any significant excess above the expected background, upper limits on the signal production cross-section for each of the signal scenarios considered are derived by using $q_{\mu}$ in the CL$_{\mathrm{s}}$ method~\cite{cls_2, cls}.  
For a given signal scenario, values of the production cross-section (parameterised by $\mu$) yielding CL$_{\mathrm{s}}<0.05$, where CL$_{\mathrm{s}}$  is computed using the asymptotic approximation~\cite{cls_3}, are excluded at $\geq$ 95\% CL.

\subsection{Likelihood fit results}\label{Results:fitResults}

The expected and observed event yields in the signal and control regions after fitting the background-only hypothesis to data, including all uncertainties, are listed in Table~\ref{Tab:FinalYields}. The total uncertainty shown in the table is the uncertainty obtained from the full fit, and is therefore not identical to the sum in quadrature of each component, due to the correlations between the fit parameters.
The compatibility of the data with the background-only hypothesis is estimated by integrating the distribution of the test statistic, approximated using the asymptotic formulae~\cite{cls_3}, above the observed value of $q_{0}$. This value is computed for each signal scenario considered, defined by the assumed mass of the heavy quark and the three decay branching ratios. The lowest $p$-value is found to be $\sim$50\%, for a $T$ mass of 700~GeV.
Thus no significant excess above the background expectation is found.

  \begin{table}[htb]{}
\begin{center}
  \caption{Event yields in the signal and control regions after the background-only fit to the signal and control regions. The uncertainties include statistical and systematic uncertainties. The uncertainties in the individual background components can be larger than the uncertainty in the sum of the backgrounds, which is strongly constrained by the data.}
    \label{Tab:FinalYields}
    \begin{tabular}{ l S[table-format=3.0, table-figures-uncertainty=1] S[table-format=3.0, table-figures-uncertainty=1] S[table-format=3.0, table-figures-uncertainty=1]} 
     \hline
     \hline
   \multicolumn{1}{l}{Sample} & \multicolumn{1}{l}{Signal region} & \multicolumn{1}{l}{Control region} \\ 
  \hline
\ttbar{}    & 39 \pm 10 & 700 \pm 70 \\
  $W$+jets   & 8 \pm 4 & 78 \pm 38 \\ 
  Single top    & 7 \pm 4 & 110 \pm 40 \\
  Others    & 10 \pm 7 & 72 \pm 48 \\
 \hline
 Total background    & 64 \pm 9 &  970 \pm 50 \\
 \hline
 Data    & 58 &  972 \\
 \hline
 \hline
 \end{tabular}
    \end{center}
  \end{table}

The sensitivity of the analysis is limited by the statistical uncertainty of the data. Including all systematic uncertainties degrades the expected mass limits by only around 20\,\GeV and for a mass of 1\,TeV, the cross-section limit increases by 4\%. 
Individual uncertainties are generally not significantly constrained by data, except for the uncertainties associated with the \ttbar{} modelling that are constrained by up to 50\% of their initial size.

A comparison of the post-fit agreement between data and prediction in the signal region, Figure~\ref{FigPostFitResult}, shows a slight deficit of data in the signal region for the $m_{T}^{\mathrm{lep}}$ distribution above 700\,\GeV.  
In this context, the observed upper limits on the $T\bar{T}$ production cross-section are slightly stronger with respect to the expected sensitivity. 
The post-fit \ttbar{} normalisation is found to be 0.93 $\pm$ 0.16 times the Monte Carlo prediction, normalised to the NNLO+NNLL cross-section.

\begin{figure}[htb!]
\centering
\includegraphics[width=0.48\textwidth]{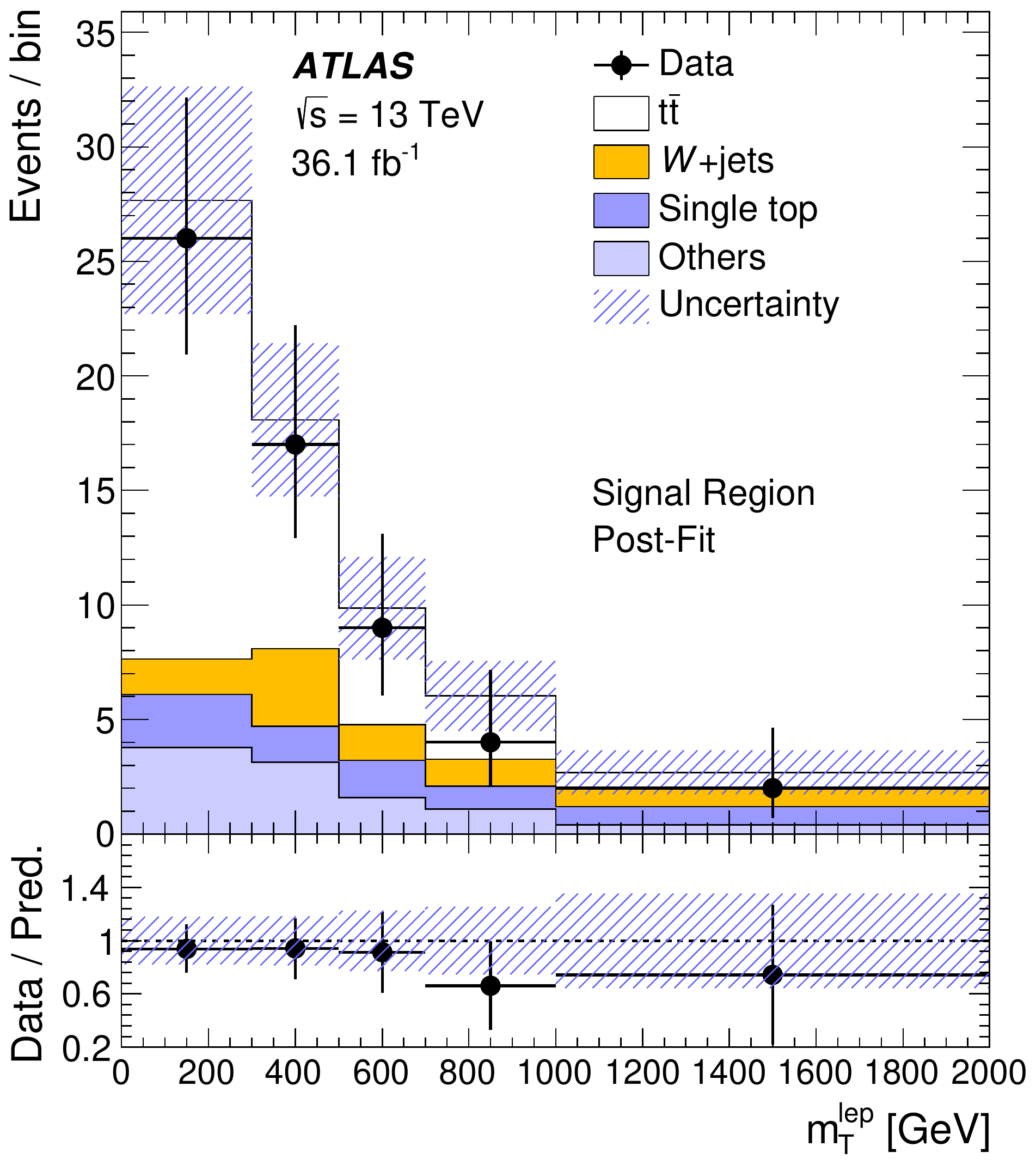} 
\includegraphics[width=0.48\textwidth]{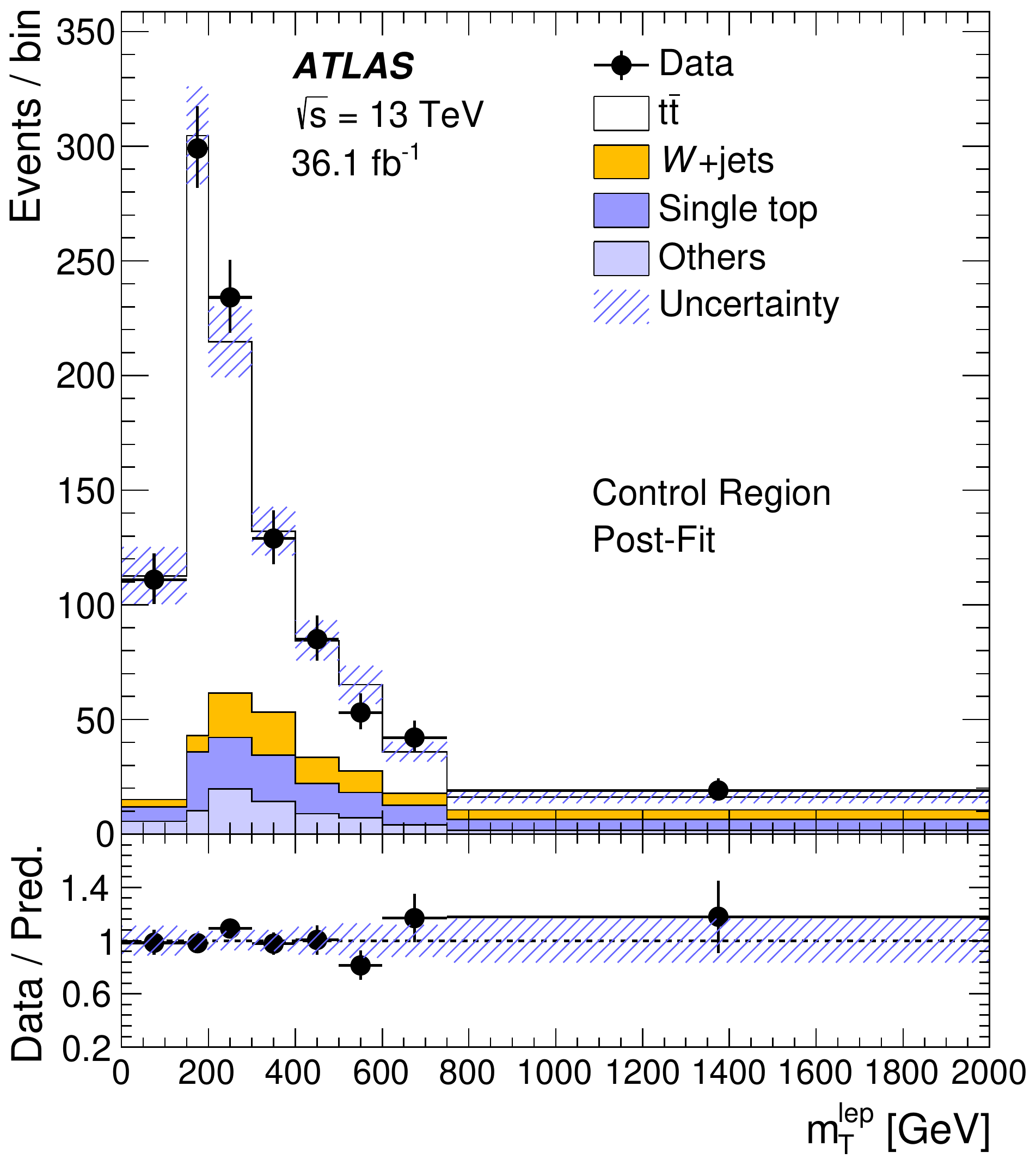} 
\caption{Fit results (background-only) for the leptonic VLQ candidate mass distributions ($m_T^{\mathrm{lep}}$) in (left) the signal region and (right) the control region. The lower panel shows the ratio of data to the fitted background yields. The band represents the systematic uncertainty after the maximum-likelihood fit.}
\label{FigPostFitResult}
\end{figure}

\subsection{Limits on VLQ pair production}\label{Results:Limits}

Upper limits at the 95\% CL on the $T\bar{T}$ production cross-section are set for two benchmark scenarios as a function of $T$ quark 
mass $m_T$ and compared to the theoretical prediction from \textsc{Top++ v2.0} (Figure~\ref{FigLimitXS}). The resulting lower limit on $m_T$ is 
determined using the central value of the theoretical cross-section prediction. 
These results are only valid for new particles of narrow width.
Assuming \BR($T \to Wb$) =1, the observed (expected) lower limit is $m_{T} = 1350$\,\GeV{} $(1310\,\GeV{})$.  For branching ratios corresponding to the SU(2) singlet $T$ scenario, the observed (expected) 95\% CL lower limit is $m_{T} = 1170$\,\GeV{} $(1080\,\GeV)$.  This represents a significant improvement compared to Run-1 searches~\cite{Aad:2015kqa,Chatrchyan:2013uxa}, for which the observed 95\% CL limit was 920\,\GeV\ when assuming \BR($T \to Wb$) =1.  

To check that the results do not depend on the weak-isospin of the $T$ quark
in the simulated signal events, a sample of \TTbar{} events with a
mass of 1.2\,\TeV{} was generated for an SU(2) doublet $T$ quark and
compared to the nominal sample of the same mass generated with an
SU(2) singlet $T$ quark.  Both the expected number of events and expected
excluded cross-section are found to be consistent between those two
samples.  Thus the limits obtained are also applicable to VLQ models with
non-zero weak-isospin.  
As there is no explicit use of charge
identification, the \BR$(T \to Wb) = 1$ limits are found to be
applicable to the pair-production of vector-like $Y$ quarks of charge
$-4/3$, which decay exclusively to $Wb$.

Exclusion limits on $T$ quark pair-production are also obtained for different values of $m_T$ and as a function of branching ratios to each of the three decays. In order to probe the complete branching-ratio plane spanned by both processes, the signal samples are weighted by the ratios of the respective branching ratios to the original branching ratios in \textsc{Protos}. Then, the complete analysis is repeated for each point in the \BR\ plane. 
Figure~\ref{FigLimitGrid} shows the corresponding expected and observed $T$ quark mass limits in the plane \BR($T \to Ht$) versus \BR($T \to Wb$), obtained by linear interpolation of the calculated CL$_s$ versus $m_{T}$.

\begin{figure}[h!]
\centering
\includegraphics[width=0.7\textwidth]{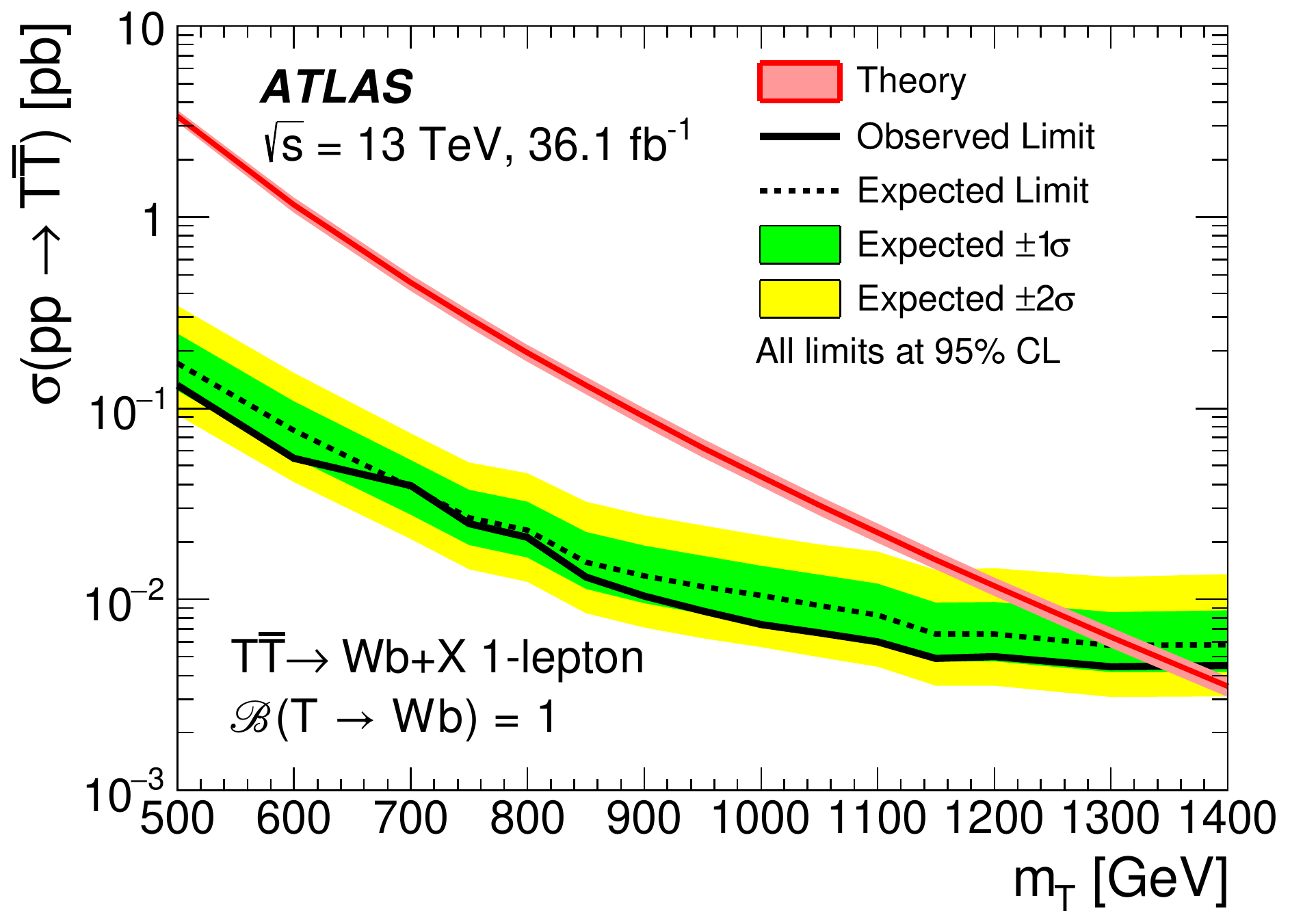} 
\includegraphics[width=0.7\textwidth]{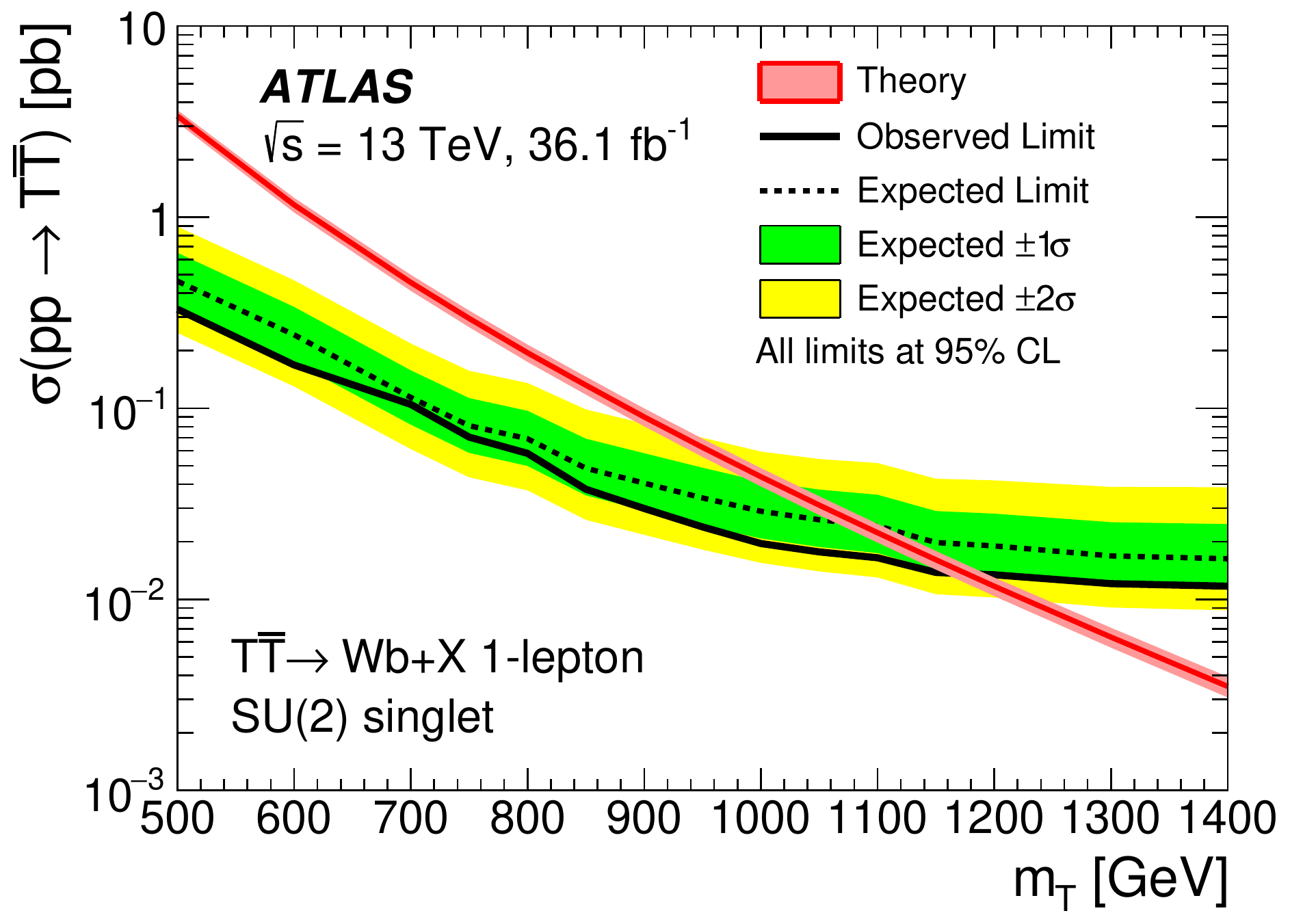} 
\caption{Expected (dashed black line) and observed (solid black line) upper limits at the 95\% CL on the $T\bar{T}$ cross-section as a function of $T$ quark mass assuming \BR$(T \to Wb) = 1$ (top) and in the SU(2) singlet $T$ scenario (bottom). The green and yellow bands correspond to $\pm$1 and $\pm$2 standard deviations around the expected limit. The thin red line and band show the theoretical prediction and its $\pm$1 standard deviation uncertainty.}
\label{FigLimitXS}
\end{figure}

\begin{figure}[h!]
\centering
\includegraphics[width=0.7\textwidth]{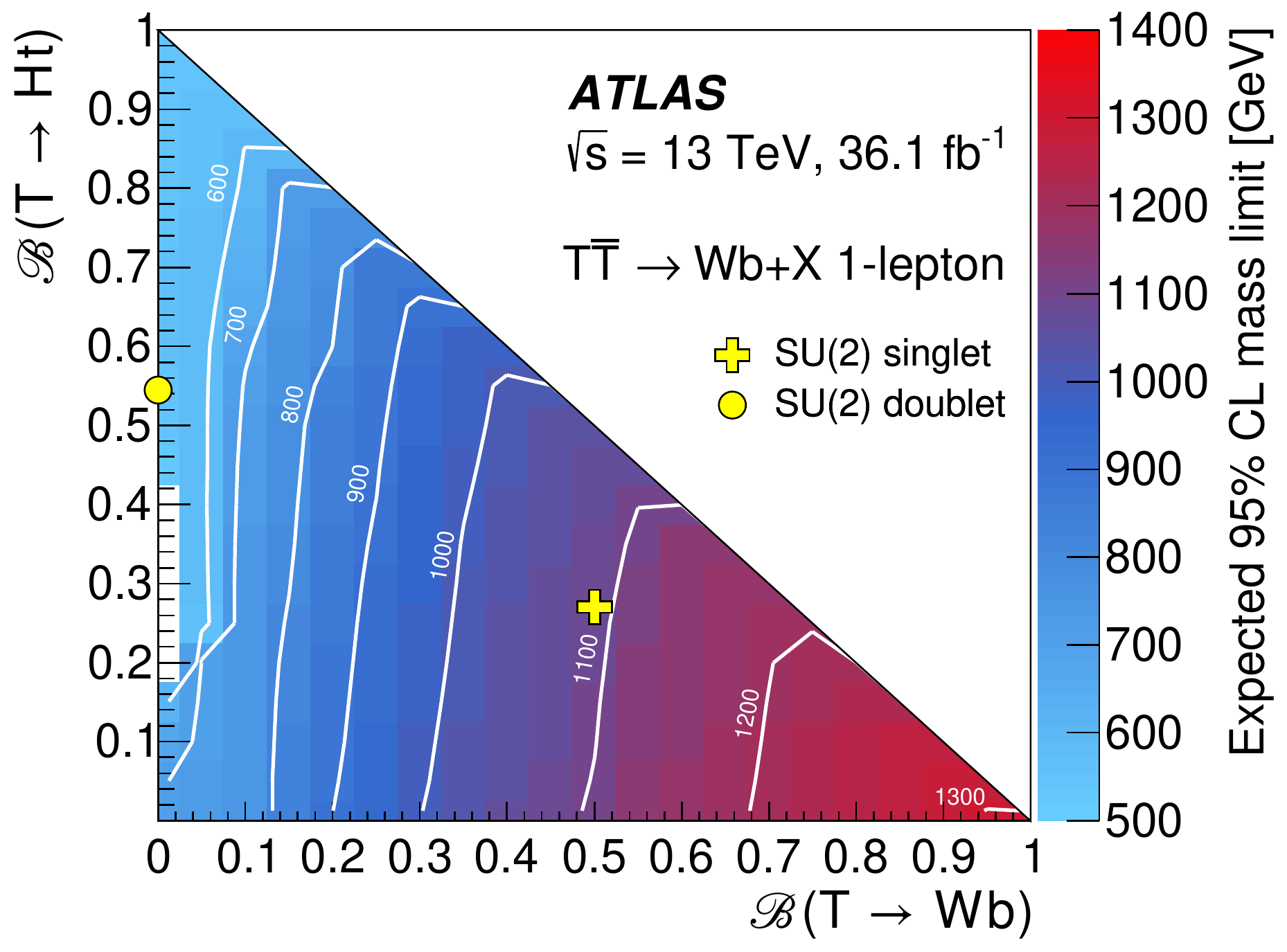} 
\includegraphics[width=0.7\textwidth]{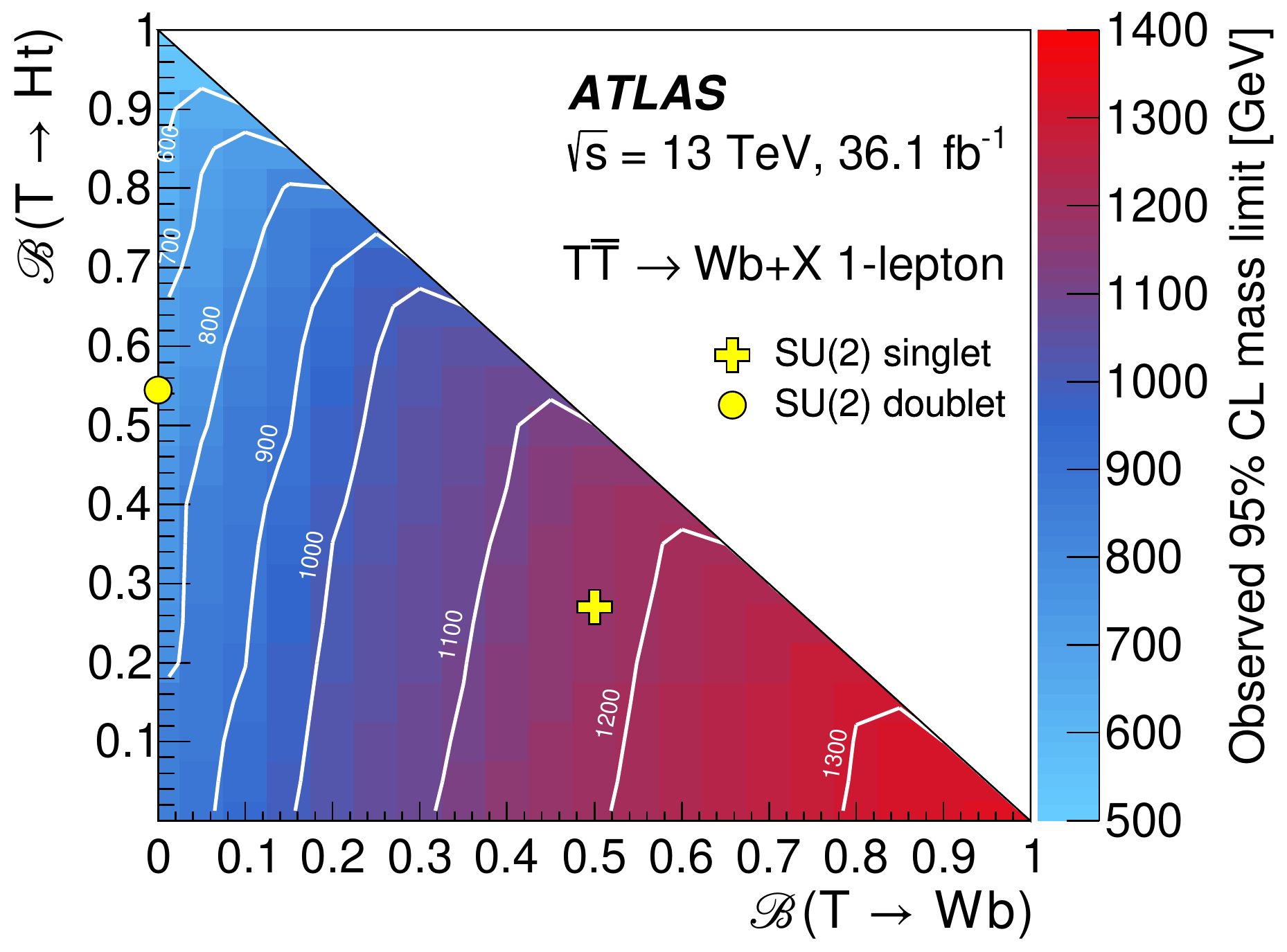} 
\caption{Expected (top) and observed (bottom) 95\% CL lower limits on the mass of the $T$ quark as a function of the decay branching ratios into \BR($T \to Wb$) and \BR($T \to Ht$). Contour lines are provided to guide the eye. The markers indicate the branching ratios for the SU(2) singlet and doublet scenarios with masses above $\sim$0.8~TeV, where they are approximately independent of the VLQ $T$ mass.  The white region is due to the limit falling below 500\,\GeV{}, the lowest simulated signal mass. }
\label{FigLimitGrid}
\end{figure}

In this search, the acceptance for VLQ \BBbar\ pair production is $\sim$3\% for the \BR$(B\to Wt) =1$ scenario and $\sim$1.3\% for the SU(2) singlet $B$ scenario, which is similar to the \TTbar\ final state. Nonetheless, the sensitivity to \BBbar\ production is expected to be weaker, as the reconstructed $T$ mass distribution is used as the final discriminant.
Without any modifications to the analysis to specifically target \BBbar\ production, observed (expected) lower limits at 95\% CL are set at 1250 (1150)~GeV when assuming \BR$(B\to Wt) =1$ and at 1080 (980)~GeV for the SU(2) singlet $B$ scenario. This represents a significant improvement compared to Run-1~\cite{PhysRevD.93.112009} and recent Run-2 searches~\cite{Sirunyan2017} when assuming \BR($B \to Wt$) =1, for which the observed 95\% CL limit was 880\,\GeV\ and 1020\,\GeV, respectively.
Being agnostic to the charge of the VLQ, the limits for \BR$(B \to Wt) = 1$ are found to be applicable to vector-like $X$ quarks of charge $+5/3$, which exclusively decay to $Wt$.
Figure~\ref{FigLimitGrid_BB} shows the corresponding expected and observed $B$ quark mass limits in the plane \BR($B \to Hb$) versus \BR($B \to Wt$), assuming $\text{\BR}(B \to Hb)+\text{\BR}(B \to Wt)+\text{\BR}(B \to Zb) =1$ .

\begin{figure}[h!]
\centering
\includegraphics[width=0.7\textwidth]{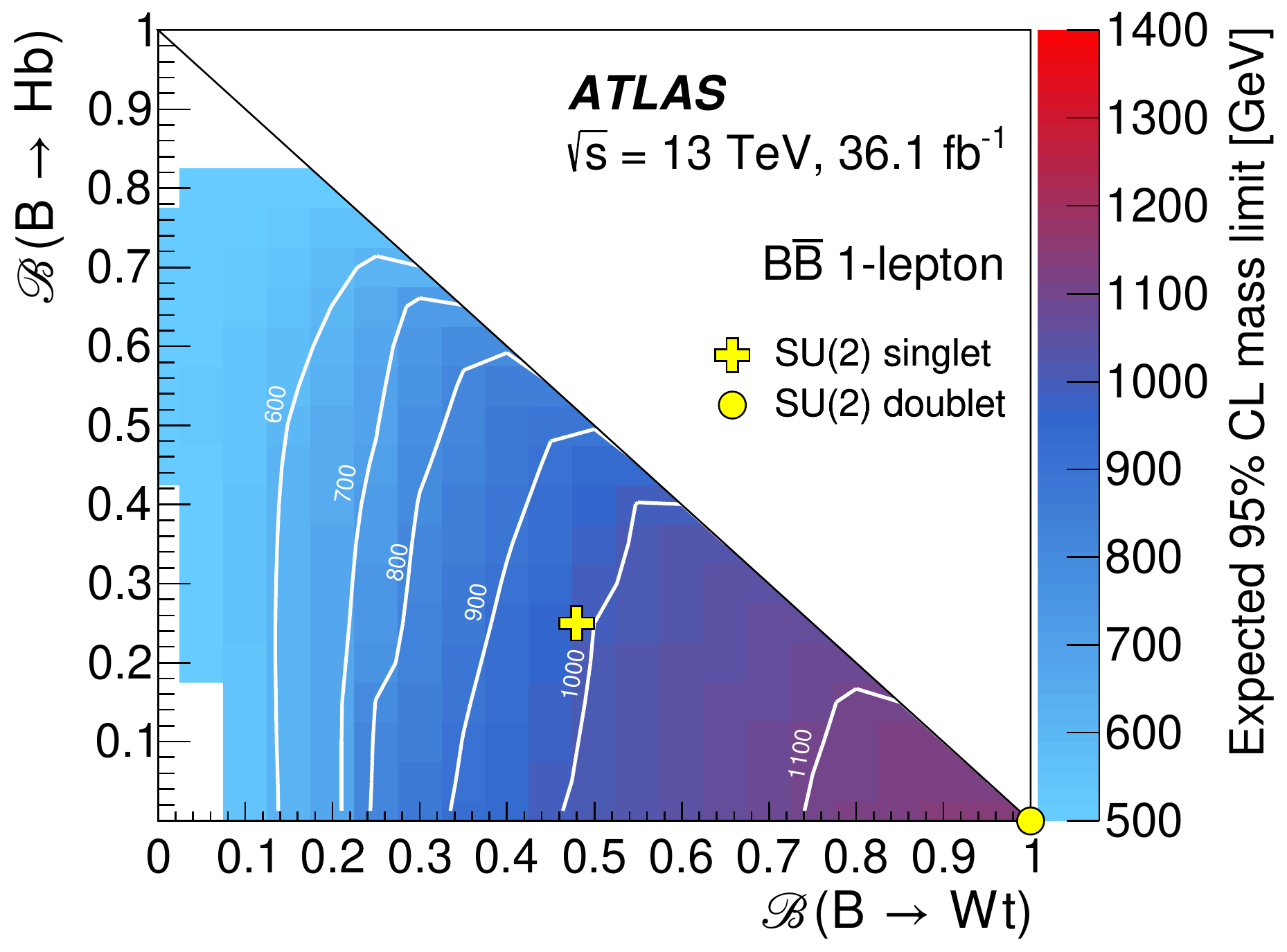} 
\includegraphics[width=0.7\textwidth]{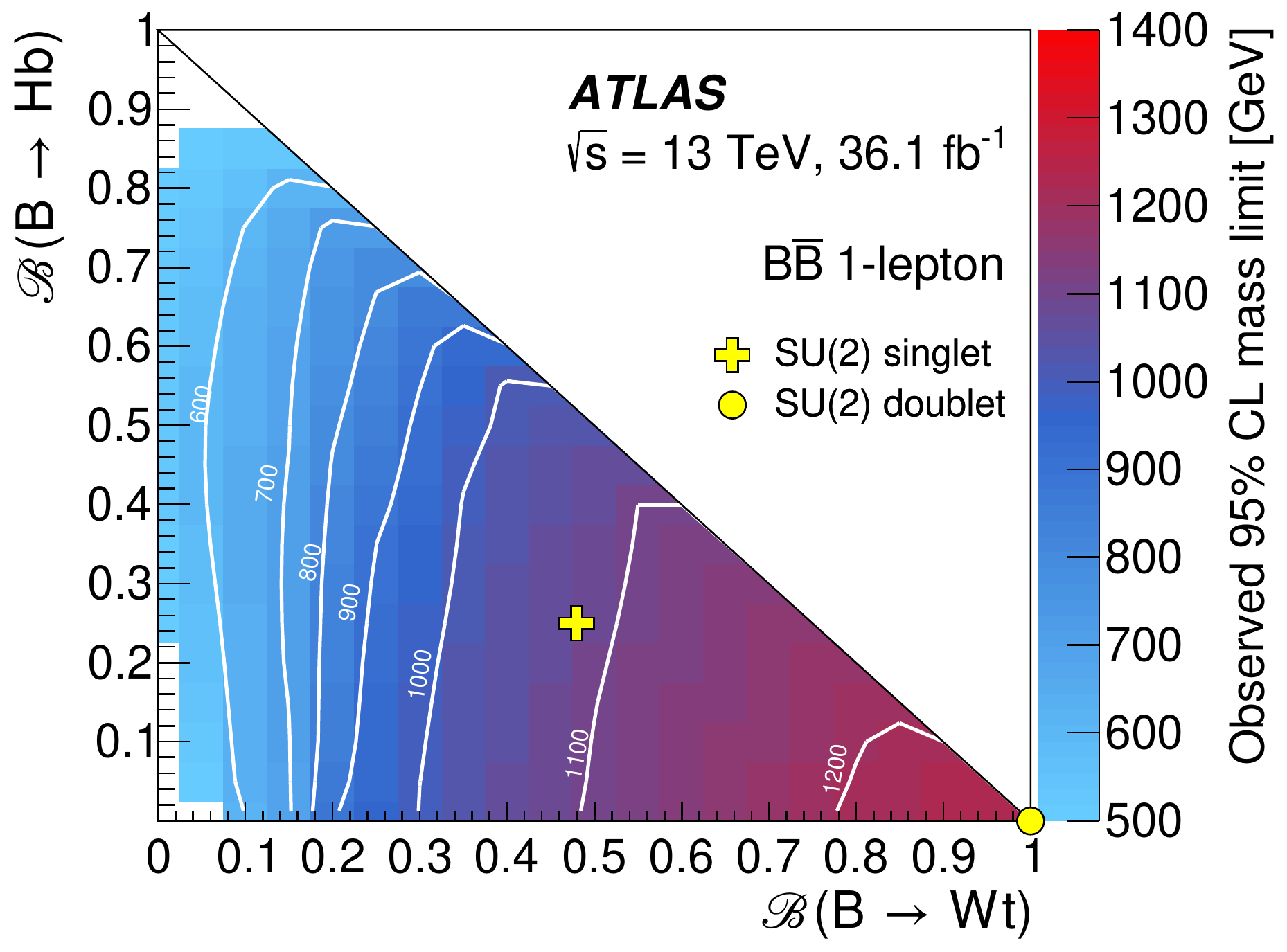} 
\caption{Expected (top) and observed (bottom) 95\% CL lower limits on the mass of the $B$ quark as a function of the decay branching ratios into \BR($B \to Wt$) and \BR($B \to Hb$). Contour lines are provided to guide the eye. The markers indicate the branching ratios for the SU(2) singlet and doublet scenarios with masses above $\sim$0.8~TeV, where they are approximately independent of the VLQ $B$ mass.  The white regions are due to the limit falling below 500\,\GeV{}, the lowest simulated signal mass.}
\label{FigLimitGrid_BB}
\end{figure}


 \FloatBarrier
 \section{Conclusions}
\label{sec:conclusion}
A search for the pair production of a heavy vector-like $T$ quark, based on $pp$ collisions at $\sqrt{s}
= 13$\,\TeV{} recorded in 2015 (3.2\,\ifb{}) and 2016 (32.9\,\ifb{})
with the ATLAS detector at the CERN Large Hadron Collider, is
presented. Data are analysed in the lepton-plus-jets final state and no significant deviation from the Standard Model
expectation is observed.  Assuming a branching ratio \BR$(T\to Wb) =
1$, the observed (expected) 95\% CL lower limit on the vector-like
quark mass is 1350\,\GeV{} (1310\,\GeV).  For the scenario of an SU(2)
singlet $T$ quark, the observed (expected) mass limit is 1170\,\GeV{}
(1080\,\GeV).  Assuming the $T$ quark can only decay to
$Wb$, $Zt$ and $Ht$, 95\% CL lower limits are derived for
various masses in the two-dimensional plane of \BR($T\to Wb$) versus
\BR($T\to Ht$).  This search is also reinterpreted to provide
limits on $B$ quark masses. These are found to be 1250\,\GeV\,(1150\,\GeV)
assuming 100\% branching ratio to $Wt$ and 1080\,\GeV\,(980\,\GeV)
under the SU(2) singlet $B$ quark scenario.  These limits are found to be equally applicable to VLQ $Y$ quark and $X$ quark production, that decay to $Wb$ and $Wt$, respectively. Mass limits are also set as a
function of the decay branching ratios \BR($T\to
Hb$) versus \BR($T\to Wt$) assuming only the $B\to Wt$, $B\to Zb$ and $B\to
Hb$ decay modes contribute.

\section*{Acknowledgements}

We thank CERN for the very successful operation of the LHC, as well as the
support staff from our institutions without whom ATLAS could not be
operated efficiently.

We acknowledge the support of ANPCyT, Argentina; YerPhI, Armenia; ARC, Australia; BMWFW and FWF, Austria; ANAS, Azerbaijan; SSTC, Belarus; CNPq and FAPESP, Brazil; NSERC, NRC and CFI, Canada; CERN; CONICYT, Chile; CAS, MOST and NSFC, China; COLCIENCIAS, Colombia; MSMT CR, MPO CR and VSC CR, Czech Republic; DNRF and DNSRC, Denmark; IN2P3-CNRS, CEA-DSM/IRFU, France; SRNSF, Georgia; BMBF, HGF, and MPG, Germany; GSRT, Greece; RGC, Hong Kong SAR, China; ISF, I-CORE and Benoziyo Center, Israel; INFN, Italy; MEXT and JSPS, Japan; CNRST, Morocco; NWO, Netherlands; RCN, Norway; MNiSW and NCN, Poland; FCT, Portugal; MNE/IFA, Romania; MES of Russia and NRC KI, Russian Federation; JINR; MESTD, Serbia; MSSR, Slovakia; ARRS and MIZ\v{S}, Slovenia; DST/NRF, South Africa; MINECO, Spain; SRC and Wallenberg Foundation, Sweden; SERI, SNSF and Cantons of Bern and Geneva, Switzerland; MOST, Taiwan; TAEK, Turkey; STFC, United Kingdom; DOE and NSF, United States of America. In addition, individual groups and members have received support from BCKDF, the Canada Council, CANARIE, CRC, Compute Canada, FQRNT, and the Ontario Innovation Trust, Canada; EPLANET, ERC, ERDF, FP7, Horizon 2020 and Marie Sk{\l}odowska-Curie Actions, European Union; Investissements d'Avenir Labex and Idex, ANR, R{\'e}gion Auvergne and Fondation Partager le Savoir, France; DFG and AvH Foundation, Germany; Herakleitos, Thales and Aristeia programmes co-financed by EU-ESF and the Greek NSRF; BSF, GIF and Minerva, Israel; BRF, Norway; CERCA Programme Generalitat de Catalunya, Generalitat Valenciana, Spain; the Royal Society and Leverhulme Trust, United Kingdom.

The crucial computing support from all WLCG partners is acknowledged gratefully, in particular from CERN, the ATLAS Tier-1 facilities at TRIUMF (Canada), NDGF (Denmark, Norway, Sweden), CC-IN2P3 (France), KIT/GridKA (Germany), INFN-CNAF (Italy), NL-T1 (Netherlands), PIC (Spain), ASGC (Taiwan), RAL (UK) and BNL (USA), the Tier-2 facilities worldwide and large non-WLCG resource providers. Major contributors of computing resources are listed in Ref.~\cite{ATL-GEN-PUB-2016-002}.

\printbibliography


\newpage 
\begin{flushleft}
{\Large The ATLAS Collaboration}

\bigskip

M.~Aaboud$^\textrm{\scriptsize 137d}$,
G.~Aad$^\textrm{\scriptsize 88}$,
B.~Abbott$^\textrm{\scriptsize 115}$,
O.~Abdinov$^\textrm{\scriptsize 12}$$^{,*}$,
B.~Abeloos$^\textrm{\scriptsize 119}$,
S.H.~Abidi$^\textrm{\scriptsize 161}$,
O.S.~AbouZeid$^\textrm{\scriptsize 139}$,
N.L.~Abraham$^\textrm{\scriptsize 151}$,
H.~Abramowicz$^\textrm{\scriptsize 155}$,
H.~Abreu$^\textrm{\scriptsize 154}$
% ...
\bigskip
\\
% ...
$^{12}$ Institute of Physics, Azerbaijan Academy of Sciences, Baku, Azerbaijan\\
% ...
$^{88}$ CPPM, Aix-Marseille Universit{\'e} and CNRS/IN2P3, Marseille, France\\
% ...
$^{115}$ Homer L. Dodge Department of Physics and Astronomy, University of Oklahoma, Norman OK, United States of America\\
% ...
$^{119}$ LAL, Univ. Paris-Sud, CNRS/IN2P3, Universit{\'e} Paris-Saclay, Orsay, France\\
% ...
$^{137}$ $^{(a)}$ Facult{\'e} des Sciences Ain Chock, R{\'e}seau Universitaire de Physique des Hautes Energies - Universit{\'e} Hassan II, Casablanca; $^{(b)}$ Centre National de l'Energie des Sciences Techniques Nucleaires, Rabat; $^{(c)}$ Facult{\'e} des Sciences Semlalia, Universit{\'e} Cadi Ayyad, LPHEA-Marrakech; $^{(d)}$ Facult{\'e} des Sciences, Universit{\'e} Mohamed Premier and LPTPM, Oujda; $^{(e)}$ Facult{\'e} des sciences, Universit{\'e} Mohammed V, Rabat, Morocco\\
% ...
$^{139}$ Santa Cruz Institute for Particle Physics, University of California Santa Cruz, Santa Cruz CA, United States of America\\
% ...
$^{151}$ Department of Physics and Astronomy, University of Sussex, Brighton, United Kingdom\\
% ...
$^{154}$ Department of Physics, Technion: Israel Institute of Technology, Haifa, Israel\\
$^{155}$ Raymond and Beverly Sackler School of Physics and Astronomy, Tel Aviv University, Tel Aviv, Israel\\
% ...
$^{161}$ Department of Physics, University of Toronto, Toronto ON, Canada\\
% ...
$^{a}$ Also at Department of Physics, King's College London, London, United Kingdom\\
$^{b}$ Also at Institute of Physics, Azerbaijan Academy of Sciences, Baku, Azerbaijan\\
$^{c}$ Also at Novosibirsk State University, Novosibirsk, Russia\\
$^{d}$ Also at TRIUMF, Vancouver BC, Canada\\
$^{e}$ Also at Department of Physics {\&} Astronomy, University of Louisville, Louisville, KY, United States of America\\
% ...
$^{*}$ Deceased
\end{flushleft}


\end{document}